\documentclass[11pt,aps,showpacs]{revtex4}
\usepackage{epsfig}

\newcommand{\beq}{\begin{equation}}
\newcommand{\eeq}{\end{equation}}
\newcommand{\beqa}{\begin{eqnarray}}
\newcommand{\eeqa}{\end{eqnarray}}
\newcommand{\bdm}{\begin{displaymath}}
\newcommand{\edm}{\end{displaymath}}

\begin{document}

\title{Dilepton production in heavy-ion collisions\\
with in-medium spectral functions of vector mesons}
\author{E. Santini$^a$, M.D. Cozma$^{a,b}$, Amand Faessler$^{a}$, C. Fuchs$^{a}$,\\
M. I. Krivoruchenko$^{a,c}$, B. Martemyanov$^{a,c}$}
\affiliation{$^a${Institut f\"{u}r Theoretische Physik$\mathrm{,}$ T\"{u}bingen
Universit\"{a}t$\mathrm{,}$ Auf der Morgenstelle 14\\ D-72076 T\"{u}bingen$\mathrm{,}$ Germany}}
\affiliation{$^b${National Institute for Physics and Nuclear 
Engineering$\mathrm{,}$
Atomistilor 407\\ 077125 Magurele-Bucharest$\mathrm{,}$ Romania}}
\affiliation{$^c${Institute for Theoretical and Experimental Physics$\mathrm{,}$ B.
Cheremushkinskaya 25\\ 117259 Moscow$\mathrm{,}$ Russia}}
\begin{abstract}
The in-medium spectral functions of $\rho$ and $\omega$ mesons and the broadening 
of nucleon resonances at finite baryon density are calculated self-consistently
by combining a resonance dominance model for the vector meson production with 
an extended vector meson dominance model. The influence of the in-medium modifications 
of the vector meson properties on the dilepton spectrum in heavy-ion collisions 
is investigated. The dilepton spectrum is generated for the C+C reaction 
at 2.0$A$ GeV and compared with recent HADES Collaboration data. The collision dynamics is 
then described by the T\"ubingen relativistic quantum molecular dynamics transport model. 
We find that an iterative 
calculation of the vector meson spectral functions that takes into account the
broadening of the nucleon resonances due to their increased in-medium 
decay branchings  is convergent and provides a reasonable 
description of the experimental data in the mass 
region $0.45\leq M \leq 0.75$ GeV. On the other side, the theoretical calculations 
slightly underestimate the region $m_\pi\leq M \leq 0.4$ GeV. Popular in-medium scenarios
such as a schematic collisional broadening and dropping vector mesons masses 
are discussed as well.
\end{abstract}


\maketitle

\section{Introduction}

It is a well-established fact that hadrons change their properties in 
a dense and excited nuclear medium. Such changes are reflected in 
mass shifts and/or in the development of complex spectral properties. 
A typical example is the nucleon that suffers a substantial mass 
shift at finite density, see, e.g., Ref.~\cite{dalen05}, but maintains its 
good quasiparticle properties. Besides a shift of the pole mass, 
resonances, both nucleonic and mesonic, have the tendency to be broadened 
and to develop spectral distributions that may even lead to a loss 
of good quasiparticle properties. For example, total photoabsorption cross sections 
on heavy nuclei \cite{Bianchi:1993nh,Krusche:2001ku} provide evidence for 
a substantial collisional broadening or melting of nucleon resonances inside 
the medium \cite{Kondratyuk94}.

To study the medium modifications of hadrons is of particular interest, 
since it not only provides insight into the properties of the strongly 
interacting hadronic many-body systems but also allows conclusions to be drawn on 
QCD ``observables'' that characterize the medium. A prominent example is 
the scalar quark condensate  $\langle {\bar q}q \rangle$ which determines 
the chiral symmetry breaking scale of QCD in the nonperturbative sector.  

Heavy-ion reactions present therefore a unique opportunity for the study 
of nuclear or hadronic matter under extreme conditions, i.e., at supranormal 
densities and high temperatures. Photo- or hadron-induced reactions 
on the nucleus provide complementary information on cold matter at moderate 
densities. 

The light vector mesons $\rho$ and $\omega$ are both 
of particular interest, because their decay into dileptons allows one to probe 
the electromagnetic 
response of the medium. For this purpose, electromagnetic probes such 
as dilepton  pairs have proven to be most efficient, 
since they leave the medium 
essentially undistorted by final-state interactions. In heavy-ion reactions, 
they provide a clear view of the effective degrees of freedom at high 
baryon density and temperature.

Theoretically, an abundance of models can 
predict the changes of vector meson masses and
widths in high density, high temperature nuclear matter, and they can be roughly 
divided in four different classes:  
Brown-Rho scaling~\cite{Bro91}, models based on QCD sum rules~
\cite{Hat92,leupold01,Zschocke:2002mp,Thomas:2005dc,kaiser97}, 
dispersion relations 
\cite{kaiser97,Kondratyuk:1998ec,Eletsky:1999mj,elka},
and  effective hadronic 
models~\cite{Her93,Klingl:1996by,Fri97,kaiser97,Urb98,rapp00,Lutz:2001mi,
Pos01,leupold04,Shklyar:2004ba,muehlich06}. 
The first approaches to the description of the in-medium vector
mesons were based on effective field theories (EFTs) \cite{Chin:1977iz}
and the Nambu--Jona-Lasinio (NJL) model \cite{Bernard:1988db}.
In some aspects, the various approaches come to qualitatively similar 
conclusions; however, the overall situation is still unclear. 
While Brown-Rho scaling, at least in its naive form, predicts a common 
downward 
mass shift of the vector mesons where the quasiparticle properties are 
essentially maintained, the hadronic models come to different conclusions. 
Concerning the $\rho$ meson, these sets of models predict in general a 
significant broadening of the  $\rho$ and the development of complex 
structures in the spectral functions, e.g., the appearance of additional 
peaks caused by the coupling to nucleon resonances. In some 
cases, this occurs in line with a slight shift of the quasiparticle peak 
which corresponds to an additional mass shift \cite{kaiser97}. 
Concerning the $\omega$ meson, the situation is even less clear. Early QCD sum 
rules calculations predicted even a repulsive mass shift \cite{Hat92}, while 
in Refs. \cite{Zschocke:2002mp,Thomas:2005dc} the strong dependence 
of the  $\omega$ properties on the higher order unknown quark condensates 
has been pointed out, which leaves room for mass shifts in both directions. 
The hadronic approaches predict in common an essential broadening of 
the   $\omega$, although they range from a strong downward mass shift 
\cite{kaiser97} to a slight upward mass shift \cite{Pos01,muehlich06} to an essential 
repulsive mass shift \cite{Lutz:2001mi}.

However, recent progress from the experimental side allows one, at least partially, 
to constrain the various theoretical 
models. While the CERES~\cite{Aga95,ceres2} and 
HELIOS~\cite{Maz94} dilepton experiments at the CERN Super Proton Synchroton (SPS) revealed clear 
evidence for in-medium effects in heavy-ion reactions (Pb+Au) through the 
observed 
enhancement of the dilepton spectra below the the $\rho$ and $\omega$ peaks 
relative to standard hadronic cocktail sources, such a behavior could
be explained either within a scenario of a 
dropping $\rho$ vector meson mass~\cite{Li:1995qm} or by the
inclusion of in-medium spectral functions for the 
vector mesons~\cite{Urb98}. Thanks to unprecedented resolution, 
the recent NA60 dimuon experiment \cite{na60} was able to ``measure''  
the in-medium $\rho$ spectral function under the conditions of 
ultrarelativistic heavy-ion collisions.  NA60 seems to rule out 
a naive dropping mass scenario but supports the picture of 
modified $\rho$-$\omega$ spectral functions predicted by 
hadronic many-body theory \cite{vanHees:2006ng}.

A second set of heavy-ion experiments
have been performed at laboratory energies of 1.0$A$ GeV (Ca+Ca and C+C) by the 
DLS
Collaboration at the LBNL Bevelac~\cite{Por97,DLS2}. Also in this case, the low mass 
region of the dilepton spectra is underestimated by present transport 
calculations, in contrast to similar measurements (1.04-4.88 GeV/nucleon)
for the $p$+$p$ and $p$+$d$ systems. 
As opposed to the ultrarelativistic case, 
the situation does not improve when the
in-medium spectral functions or the dropping mass scenarios 
are taken into account~\cite{BCRW98,Ern98} (the DLS
puzzle). Other scenarios such as possible contributions from the 
quark-gluon plasma or in-medium modifications
of the $\eta$ mass have been excluded as a possible resolution 
of this puzzle. Decoherence effects~\cite{She03}
have proven to be partially successful in explaining the difference 
between the DLS data and the theoretical predictions. 
However, in this energy regime, which probes the high density, low temperature 
phase, the situation is going to be improved significantly with the 
already existing and forthcoming measurements of 
the HADES Collaboration at GSI \cite{Had05,Had06,Frohlich:2006cz}.
Complemented are the heavy-ion experiments by  $\gamma$-nucleus reactions. 
The CB-TAPS experiment \cite{cbelsa}, which focused exclusively on 
the $\omega$ meson 
and reported an enhanced strength below the $\omega$ peak, reports a 
broadening of the   $\omega$ observed in $\gamma$-nucleus reactions. Also, 
the dilepton mass spectrum measured at Japan's National Laboratory for High Energy Physics 
(KEK) in $p+A$ reactions at a beam
energy of 12 GeV \cite{KEK,KEK2} revealed an excess of the dileptons 
below the $\rho $-meson peak over known sources. However, these data could not be 
explained within the standard dropping mass scenario and/or 
assuming a significant collision broadening of the vector mesons
\cite{Elena}. An enhanced bremsstrahlung contribution, which is 
presently under debate at low energies \cite{Kaptari:2005qz,Bratkovskaya:2007jk}, 
will most likely not help explain the high energy KEK data.

A major difficulty in the interpretation of heavy-ion collision experimental 
data lies in the fact that the gap between observables and theoretically predicted 
in-medium properties of hadrons has to be filled by transport models. Transport 
models account for complicated reaction dynamics and provide the link between
theory and experiment. A drawback of such a procedure is dependence on the phenomenology 
and an extended set of input parameters entering the models. Usually, a significant
fraction of such parameters can be neither  constrained by data nor based on 
well-established theoretical approaches. This certainly diminishes the possibility 
of testing experimentally theory and drawing physical conclusions from 
the experimental data.

In the present work, we remove some model uncertainties by applying 
a {\it unified} description of vector meson production, vacuum decays, and 
in-medium properties of vector mesons. For this purpose, we use a resonance dominance model 
for nucleon-nucleon scattering in combination with an extended vector meson 
dominance (eVMD) model. Nucleon resonance dominance (NRD) is 
an effective principle which assumes that vector meson production runs 
over the excitation of nucleon resonances \cite{Pos01,krivo02,resdec,crosssec}. 
On the other hand, eVMD introduces radially excited $\rho $ and $\omega $ mesons 
\cite{Fae00} in the $RN\gamma $ transition form factors \cite{krivo02} in order 
to fulfill the quark counting rules as a strict consequence of QCD \cite{Matveev:1973ra}. 
This allows the kinematically complete, gauge invariant, fully relativistic, and unified description 
of the nucleon resonance transition amplitudes $R\rightarrow NV$ ($V=\omega ,\rho $), $R\rightarrow
N\gamma $, $\gamma ^{*}N\rightarrow R$ (electro-production), and 
$R\rightarrow Ne^{+}e^{-}$ with arbitrary spin and parity in terms of
the magnetic, electric, and Coulomb transition form factors. 
The eVMD model solves a long-standing problem of VMD which underestimates the 
$\rho$-meson branchings of nucleon resonances when the normalization to 
the photon branchings is performed. The parameters of eVMD are fixed by fitting to 
photoproducion and electroproduction experimental data, by using results of 
the $\pi N$ multichannel partial-wave analysis, and, when the experimental 
data are not available, by using predictions of the quark models \cite{krivo02}.

Once the model parameters are fixed, one obtains a unified (and parameter free) 
description of quite a broad range of physical processes including vector meson 
decays, nucleon resonance decays to vector mesons and dileptons, and vector meson 
and dilepton production in elementary and heavy-ion reactions. The NRD+eVMD model 
has successfully been applied earlier to vector meson ($\omega$ and $\phi$) production 
in elementary ($p+p$) reactions \cite{omega,phi} and dilepton production in elementary 
$p+p$ and $p+d$ reactions \cite{resdec}. In Ref. \cite{omega04} it has been further
demonstrated that this model is qualitatively able to explain the $\omega$ and $\phi$ 
angular distributions in $p$+$p$ reactions \cite{DISTO,COSYTOF}. 

Embedded within the framework of the T\"ubingen relativistic quantum molecular 
dynamics (RQMD) transport model \cite{Uma98,fuchs05}, the NRD+eVMD model has been 
applied to heavy-ion reactions without introducing new parameters \cite{She03,Cozma06}. 
The comparison with the dilepton data from DLS \cite{Por97,DLS2} and HADES \cite{Had06} 
collaborations revealed clear evidence for the in-medium effects required, in particular, 
to suppress excessive dilepton production from the $\omega$-meson decays. In Refs. \cite{She03,Cozma06} 
the collisional broadening and dropping the vector meson masses have been analyzed 
phenomenologically. In the present work, we go beyond the phenomenological analysis 
by calculating the in-medium spectral functions of the $\rho$ and 
$\omega$ mesons and nucleon resonances using the NRD+eVMD approach. This allows 
the first self-consistent theoretical description to be made of dilepton spectra 
based on a unified model for nucleon resonances, vector mesons, and dilepton production, 
and their in-medium modifications.

\section{In-medium spectral functions}
\subsection{Resonance model}
The in-medium properties of hadrons are generally expressed in 
terms of the self-energy  $\Sigma_V$. The self-energy determines the 
spectral function of the quasiparticle in the medium. As long as the 
self-energy shows only a moderate energy dependence, 
the real part of  $\Sigma_V$ can be interpreted in terms of a mass shift, 
while the imaginary part generates the in-medium width. To leading 
order in density, the  self-energy is determined by the forward scattering 
length of the hadron with the surrounding particles. Since the 
$\rho$-nucleon and $\omega$-nucleon scattering lengths are unknown from 
the experimental side, these quantities have to be determined theoretically.

In the present work, we apply the resonance model to calculate 
the forward scattering of vector mesons on nucleons. The resonance 
model is not a field theory in the strict sense where 
corresponding Feynman diagrams are evaluated but rather an effective model 
that has some similarity to a field theory based on Feynman diagrams 
with the intermediate resonances in the $s$ channel of vector meson
and nucleon scattering. Such an approach was applied in many 
previous investigations of vector mesons properties in the 
nuclear medium \cite{Pos01,elka,Penner02,leupold04,cabrera}.
The present approach differs with respect to previous investigations 
by the fact that in the NRD+eVMD model the corresponding couplings of 
resonances to the nucleon and vector meson are of relativistic form and 
kinematically complete.

The self-energy $\Sigma_V$ of a vector meson $V$ in an isotopically symmetric 
nuclear medium is determined by the 
invariant $VN$  forward scattering amplitude $A_{VN}$ 
\beq
\Sigma_V = -\int A_{VN}~ 2\times 2 \frac{d^3p_N}{2E_N(2\pi)^3}~.
\label{sigma}
\eeq
Here $V$ refers either to a $\rho^0$ or a $\omega$  meson. Due to isosymmetry
of the medium, the self-energy $\Sigma_V$ for $\rho^\pm$ mesons is
the same as for $\rho^0$ meson. The forward scattering amplitude $A_{VN}$
is the same for proton ($N=p$) and neutron ($N=n$) scattering. 
The integral in Eq.~(\ref{sigma}) runs over the nucleon momenta within 
the Fermi sphere with Fermi momentum determined by
nuclear matter density $\rho_B$
\beq
\rho_B =\frac{ 2}{3\pi^2}p_F^3~.\label{n}
\eeq
The amplitude $A_{VN}$ is of Breit-Wigner form for resonance scattering
\beq
A_{VN} = - \sum_{R}\frac{(2J_R+1)}{2\times 3} \frac{8\pi s}{k}
\frac{\Gamma_{RNV}(s)}{s-M_R^2 +i\sqrt{s}\Gamma_R^{\mathrm{tot}}(s)}~.
\label{ampl1}
\eeq
In Eq.~(\ref{ampl1}) the scattered vector meson has 
running mass squared $M^2$ and momentum
$p$, $s=(p_N+p)^2$ is the running mass squared of the baryon resonance $R$,
and $k$ is the  c.m. momentum. The width $\Gamma_{RNV}(s)$ refers to the
decay of the baryon resonance $R$ to nucleon $N$ and vector meson $V$ with
fixed mass squared $M^2$. 

The width $\Gamma_R^{\mathrm{tot}}(s)$ refers to the
decays of resonance $R$ not modified by the medium, in particular, with the
vacuum spectral functions for the decay products. This represents the first
approximation in the calculation of the medium contribution $\Sigma_V$ to 
the total self-energy $\Sigma_V^{\mathrm{tot}} =\Sigma_V +\Sigma_V^{(0)} $ of the vector 
meson $V$. The vacuum self-energy  $\Sigma_V^{(0)}$ is determined by the corresponding 
 vacuum width
\beq
\Im\Sigma_V^{(0)} = - m_V \Gamma_V^{\mathrm{tot}}(M),~ \Re\Sigma_V^{(0)}=0~.
\label{Sigma0} 
\eeq
Here $\Gamma_{\rho}^{\mathrm{tot}}(M)$, $\Gamma_{\omega}^{\mathrm{tot}}(M)$ are 
essentially given by the decay widths of the $\rho$ meson into two pions 
and of the $\omega$ meson into three pions, respectively. The two-pion decay 
width of the $\rho$ meson is given by
\beq
\Gamma_{\rho}^{\mathrm{tot}}(M)=\Gamma_{\rho}^{\mathrm{tot}}(m_{\rho})\,\,\frac{m_{\rho}}{M}\,
\left(\frac{k_{\pi}(M,m_{\pi},m_{\pi})}{k_{\pi}(m_{\rho},m_{\pi},m_{\pi})}
\right)^3
\Theta(M^2-4m_{\pi}^2)\\
\label{gamrho}
\eeq
where $k_{\pi}(M,m_{\pi},m_{\pi})$ is the momentum of the pions in the 
rest frame of the decaying $\rho$ meson having mass $M$; 
$m_{\rho}$ is the physical $\rho$ meson mass 
and $\Gamma_{\rho}^{\mathrm{tot}}(m_{\rho})$~=~150 MeV the on-shell decay width.
The three-pion decay width of the $\omega$ meson can be calculated 
according to the two-step process $\omega \rightarrow\rho\pi \rightarrow 3\pi$ 
as proposed by Gell-Mann, Sharp, and Wagner \cite{gom}. 
The corresponding result can be parametrized in the simple form
\beq
\Gamma_{\omega}^{\mathrm{tot}}(M)=
\Gamma_{\omega}^{\mathrm{tot}}(m_{\omega})\,\,\frac{m_{\omega}}{M}\,
\left(\frac{M^2-9m_{\pi}^2}{m_{\omega}^2-9m_{\pi}^2}\right)^3
\Theta(M^2-9m_{\pi}^2)
\label{gamomega}
\eeq
with $m_{\omega}$ the physical $\omega$-meson mass, 
and $\Gamma_{\omega}^{\mathrm{tot}}(m_{\omega})$~=~8.4 MeV the on-shell decay width.

In the next order, the medium modification of the resonance 
spectral function including the modification of the resonance width
due to the modifications of products of the resonance decay should be taken 
into account.

The width $\Gamma_{RNV}(s)$ can be expressed by the helicity amplitudes
$A_\frac32 = <1-\frac12|S|\frac32>$, $A_\frac12 = <1\frac12|S|\frac12>$, 
$S_\frac12 = <0-\frac12|S|\frac12>$ 
of the $R\rightarrow NV$ decay \cite{krivo02}

\beq
\Gamma_{RNV}(s) = \frac{k}{8\pi s}\frac{2(A_\frac32^2+ 
A_\frac12^2+ S_\frac12^2)}
{(2J_R+1)}~.\label{GRNV}
\eeq

The calculation of these amplitudes uses the coupling constants of vector 
mesons
to the $ R N$ transition current. They were obtained in Ref.~\cite{krivo02} by 
fitting
photoproduction and electroproduction amplitudes of baryonic resonances in the eVMD 
model.
The transverse and longitudinal self-energies $\Sigma_V^T$ and
$\Sigma_V^L$ can be obtained by the following substitutions in Eq. (\ref{GRNV}):
\beqa
&&\frac23(A_\frac32^2+ A_\frac12^2+ S_\frac12^2)\rightarrow 
(A_\frac32^2+ A_\frac12^2)
\frac{1+\cos^2\theta}{2}+2S_\frac12^2\frac{\sin^2\theta}{2}~,\\
&&\frac23(A_\frac32^2+ A_\frac12^2+ S_\frac12^2)\rightarrow 
2S_\frac12^2\cos^2\theta
+(A_\frac32^2+ A_\frac12^2)\sin^2\theta~,
\eeqa
where $\theta$ is the polar angle of vector meson momentum in the
c.m. system. The polarization averaged self-energy $\Sigma_V$ reads then 
\beq
\Sigma_V=\frac{2\Sigma_V^T+\Sigma_V^L}{3}~.
\eeq
The vector meson spectral function ${\cal A}_V$ is defined by the off-shell 
self-energy $\Sigma_V^{\mathrm{tot}} (M,|{\bf p}|)$ as follows
\beq
{\cal A}_V (M,|{\bf p}|)=\frac{1}{\pi}\frac{-\Im\Sigma_V^{\mathrm{tot}}}
{(M^2-m_V^2-\Re\Sigma_V^{\mathrm{tot}})^2+(\Im\Sigma_V^{\mathrm{tot}})^2}~.
\label{specfun}
\eeq

The helicity amplitudes entering into Eq.~(\ref{GRNV}) have been calculated within 
the same relativistic approach \cite{krivo02} and with the same set of 
baryonic resonances $R$ that has successfully been applied to 
dilepton and vector meson production in $p$+$p$ 
collisions \cite{resdec,omega,phi}.
This includes the following set of resonances for $\rho N$  and  
$\omega N$ scattering: 
$N^*(1535)\frac{1}{2}^-,~N^*(1650)\frac{1}{2}^-,~N^*(1520)\frac{3}{2}^-,
~N^*(1440)\frac{1}{2}^+,~N^*(1720)\frac{3}{2}^+,~
N^*(1680)\frac{5}{2}^+,$
$\Delta(1620)\frac{1}{2}^-,~\Delta(1700)\frac{3}{2}^-,
~\Delta(1232)\frac{3}{2}^+,~\Delta(1905)\frac{5}{2}^+,
~\Delta(1950)\frac{7}{2}^+$.

A straightforward extension of the approach to finite temperature and 
baryon chemical potential would be to integrate the present amplitudes, 
Eq. (\ref{GRNV}), 
over hot Fermi distributions. This can easily be done and will be a first step 
toward an application, e.g., at SPS conditions.  However, for a meaningful 
determination of spectral functions at SPS conditions, one would have 
to take into account the coupling to not only baryonic but also mesonic 
excitations ($ \pi,K,\dots$)  \cite{vanHees:2006ng}.

So far, analyticity has  not been used in the data analyses to 
determine the multichannel $\pi N$ scattering amplitudes 
\cite{Manley:1992yb,Feuster:1997pq,Vrana:1999nt}. The current phenomenological 
schemes provide resonance masses and widths, based on multichannel 
unitarity and other, less fundamental constraints. The background phases 
entering 
the dispersion relations are not provided.

We did not attempt to embed analyticity and restricted our approach to 
energies $s < 4$ GeV$^{2}$, where the sum over Breit-Wigner poles 
gives typically a good approximation for the amplitudes \cite{Collins:1977jy}. 
The background is described by $t$-channel $\sigma$-meson exchange and the 
$u$-channel 
part of the Compton $\rho N$
scattering diagram evaluated in the Born approximation.

\subsection{Nonresonant contributions}
 
Up to now we have not discussed possible nonresonant contributions to the 
forward vector meson-nucleon scattering. The reason is twofold. 
First, we cannot fix the nonresonant amplitudes with the same 
accuracy as the resonant ones.
Second, if we fix them with the available accuracy we 
would find that nonresonant amplitudes
approximately cancel in the sum. For example, in the case of the $\rho$ meson,
there exist the Compton scattering amplitude which gives a positive contribution
to the real part of the $\rho$-meson self-energy and the 
amplitude due to $\sigma$-meson exchange which gives a 
negative contribution to it (the latter is of the
same origin as the attractive part of the $NN$ interaction \cite{bonn}). The unknown
$\rho\rho\sigma$ coupling constant can be extracted from the width of the 
$\rho^0\rightarrow \pi^+\pi^-\pi^+\pi^-$ decay if one assumes that this
decay goes over an intermediate $\rho^0\sigma$ state.

\begin{figure}[!htb]
\centering
\includegraphics[width=.6\textwidth]{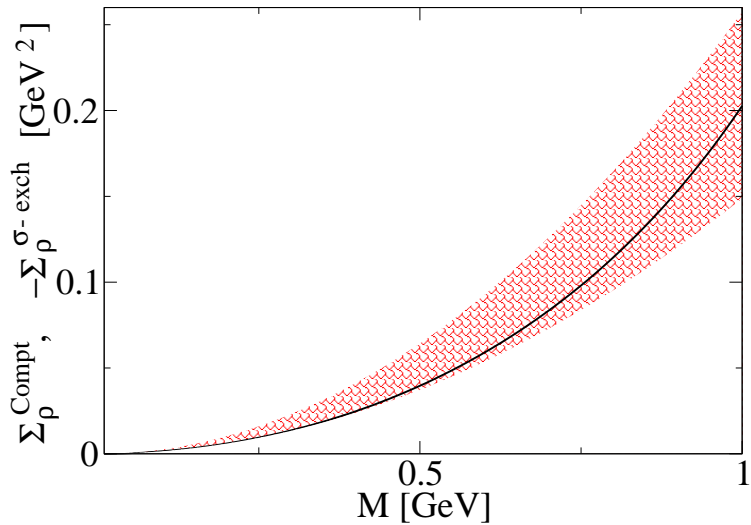}
\caption{(Color online) Nonresonant contributions to $\rho$-meson self-energy from
Compton scattering amplitude (solid line) and from the amplitude due to exchange
by $\sigma$ meson (shaded region). The shaded region corresponds to the error
in the branching ratio $\mathrm{Br}( \rho^0\rightarrow \pi^+\pi^-\pi^+\pi^-) = (1.8 \pm 0.9)\times10^{-5}$.}
\label{rhomesselfen}
\end{figure}

The two contributions from Compton scattering ($\Sigma^{\rm Compt}$) 
and $\sigma$ exchange  ($\Sigma^{\sigma-{\rm exch}}$) are shown
in Fig. \ref{rhomesselfen}. For the estimate shown in Fig.~\ref{rhomesselfen} 
the corresponding $NN\rho$  tensor 
coupling and $NN\sigma$ coupling strength were 
taken from the Bonn one-boson-exchange model \cite{bonn} for 
nucleon-nucleon scattering [$f_{NN\rho}=19.8$ (tensor coupling) and $g_{NN\sigma}=10$ ].  
The error band for $\Sigma^{\sigma-{\rm exch}}$ is due to the relatively 
large uncertainty in the four-$\pi$ decay of the $\rho$ meson
$$\mathrm{Br}( \rho^0\rightarrow \pi^+\pi^-\pi^+\pi^-) = (1.8 \pm 0.9)\times10^{-5}~.$$ 
Nevertheless, from Fig.~\ref{rhomesselfen}, one sees that the 
contributions from Compton scattering and $\sigma$-exchange   are of different sign 
and comparable magnitude. For the mean value of the 
$\mathrm{Br}( \rho^0\rightarrow \pi^+\pi^-\pi^+\pi^-) $ 
branching, they almost cancel completely, and changes of the  $\rho$-meson 
spectral function shown below are insignificant.   

To account for nonresonant contributions to the $\omega$ spectral 
function within the present scheme, we assume an $\omega\omega\sigma$ coupling 
three times larger than that for $\rho\rho\sigma$ which is motivated 
by the comparison with the two-pion coupling. The $NN\omega$ vector coupling 
($g_{NN\omega}=15.9$) is again taken from the Bonn potential  \cite{bonn}.  
As can be seen in Fig. \ref{omegamesspecfun}, the influence of the 
nonresonant contributions is now more pronounced than in the case of 
the $\rho$ meson; however, the qualitative features of the spectral 
distributions are not changed.  

The Breit-Wigner amplitudes decrease as $1/s$ with increasing $s$. 
Such a parametrization ensures
the change of the resonance phases by $\pi$ from low to high energies. 
The $\sigma$-meson exchange generates the scalar mean field, 
which is known to be important in the modification of the
nucleon masses \cite{dalen05}. It plays an important role in our scheme too. 
The component of the amplitude connected to
the $\sigma$-meson exchange remains constant for $s \to \infty$.

\subsection{$\rho$-meson spectral function}
\begin{figure}[!htb]
\centering
\includegraphics[width=.8\textwidth]{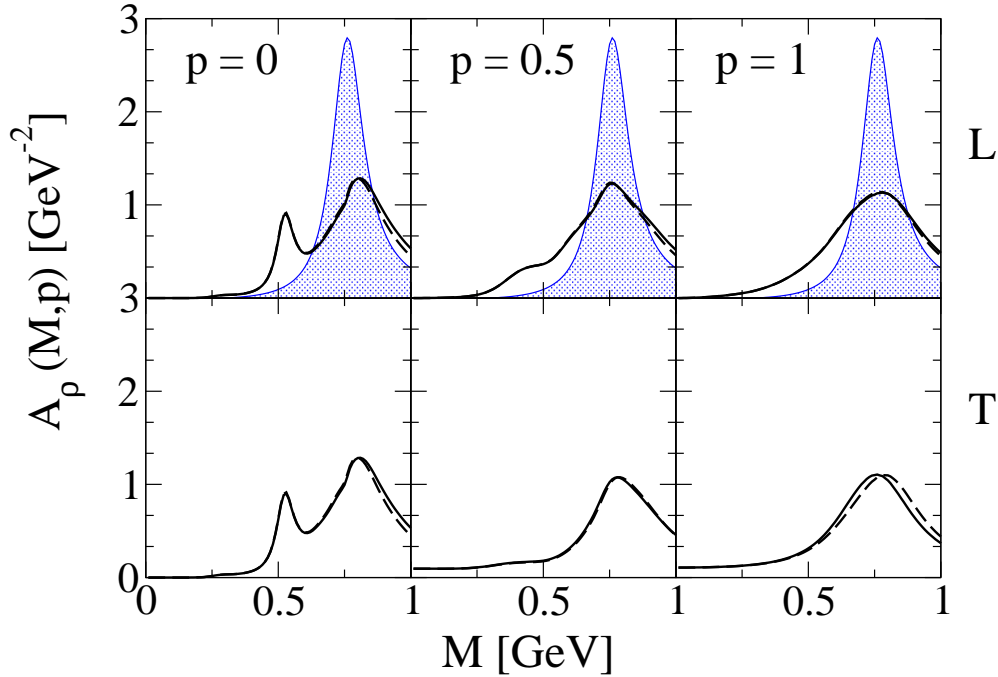}
\caption{(Color online) Longitudinal ($L$) and transverse ($T$) $\rho$ spectral functions 
in nuclear matter at saturation density for various momenta $p$ (in
GeV). Dashed lines stand for the resonance approximation, solid lines 
represent calculations that also included the nonresonant contributions. 
The shaded area shows the vacuum
spectral function.}
\label{rhomesspecfun}
\end{figure}

In the following, we discuss first the $\rho$ meson. Figure~\ref{rhomesspecfun} shows 
the $\rho$ spectral function in nuclear matter at nuclear saturation density 
$\rho_0 = 0.16~{\rm fm}^{-3}$. Longitudinal (${\cal A}^L$) and transverse (${\cal A}^T$) 
spectral functions are found to be rather similar. This 
means that unpolarized spectral functions can be used in
the calculations of dilepton spectra.

We observe 
a slight upward mass shift of the  $\rho$ and a substantial broadening. 
At low momenta, the spectral functions show a clear two-peak 
structure which vanishes with increasing vector meson momentum. 
The results shown in Fig.~\ref{rhomesspecfun} are in qualitative and even 
quantitative agreement with 
previous calculations based on the resonance model assumption 
\cite{leupold04}. Although the various approaches are based on 
different ways to describe the corresponding transition form factors, 
eVMD in the present case, and parameters are partially fixed in 
different way, this fact demonstrates the stability of the essential 
features predicted by these types of models.

The emerging two-peak structure can be understood as follows. 
The value and sign of the self-energy 
$\Re\Sigma_V$ depend on the pole positions of the particular resonances.
 If the vector meson mass squared is small, 
the invariant mass of vector meson plus nucleon is below  the pole masses
of the relevant nucleon resonances. Therefore the real part of the
vector meson self-energy is negative. This is a typical example for 
level repulsion (vector meson plus nucleon
and nucleon resonance). Consequently,  the factor
$(m^2-m_V^2-\Re\Sigma_V)^2$ in the denominator of the vector meson
spectral function, Eq. (\ref{specfun}), is small or even equal to zero. Thus
the first peak in the spectral function emerges at a vector meson mass
around $0.5 ~ \rm {GeV}$. The major contribution, which generates the 
first peak, comes from the $N^*(1520)$, which is in agreement with the 
findings reported in Ref. \cite{leupold04}.

If the vector meson mass squared lies in the vicinity of 
its vacuum value $m_V^2$, the invariant mass of vector meson plus nucleon lies 
above  the pole masses of the relevant nucleon resonances and the real part 
of the
vector meson self-energy is positive. Thus
we obtain the second peak in the spectral function at a vector meson mass
slightly above $m_V$.

  At high vector meson momenta, the invariant mass of the vector meson 
plus nucleon
 is always above  the pole masses of the relevant nucleon resonances.
  As a result, the spectral function has only one single peak
 slightly above $m_V$.

\begin{figure}[!htb]
\centering
\includegraphics[width=.6\textwidth]{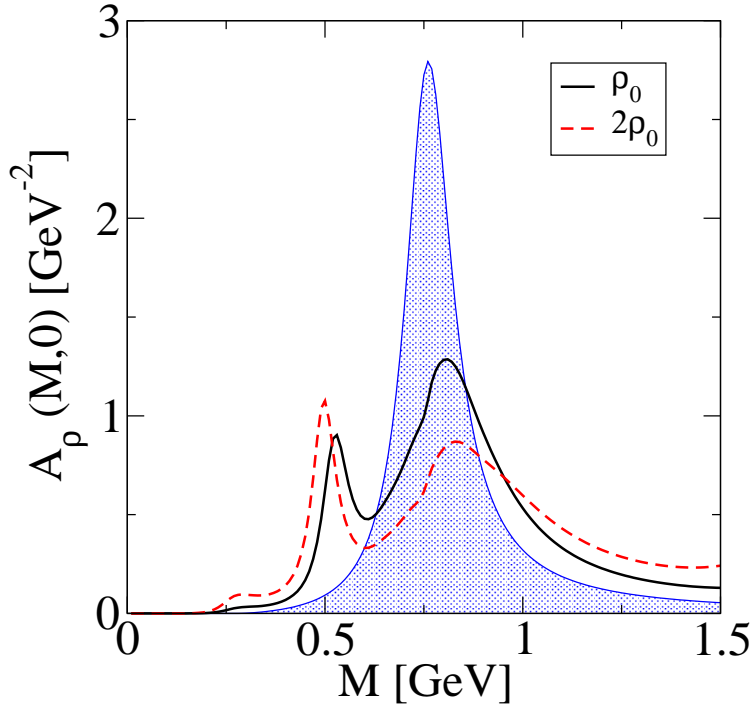}
\caption{(Color online) Unpolarized $\rho$ meson spectral function at rest
in nuclear matter at saturation density and twice saturation density. 
The shaded area displays the vacuum spectral function. 
}
\label{rhomesspecfun1}
\end{figure}
Figure \ref{rhomesspecfun1} displays finally the dependence of  the  
$\rho$-meson 
spectral function on nuclear density. It shows the 
unpolarized $\rho$-meson spectral function at rest at $\rho_0$ and at $2\rho_0$ 
nuclear density. With increasing density, we observe a further shift of 
strength 
away from the original pole mass; i.e., the first branch in spectral 
distribution 
is slightly enhanced and even shifted to lower masses, while the second peak 
is slightly shifted upward at  $2\rho_0$ compared to  $\rho_0$ and also 
additionally 
broadened. 

In this context, it should be noted that the resonance model predictions 
stand in contrast to the EFT coupled-channel calculations of 
Ref. \cite{Lutz:2001mi}, 
which predict no significant medium dependence of the $\rho$, 
concerning neither 
a mass shift nor a broadening. The reason that in the approach 
of Ref.~\cite{Lutz:2001mi} 
much less strength is shifted to lower masses lies mainly in the 
much weaker coupling to the  $N^*(1520)$ found in Ref.~\cite{Lutz:2001mi}. For 
this resonance, the value of $\Gamma_{N\rho}\sim 2$ MeV  \cite{Lutz:2001mi} 
has to be compared with  $\Gamma_{N\rho}\sim 25$ MeV 
from Refs.~\cite{krivo02,leupold04}. 
The latter value, however, agrees with that of the 
PDG \cite{PDG96} and the Manley and Saleski 
analysis \cite{Manley:1992yb}.

\subsection{$\omega$-meson spectral function}

\begin{figure}[!htb]
\centering
\includegraphics[width=.8\textwidth]{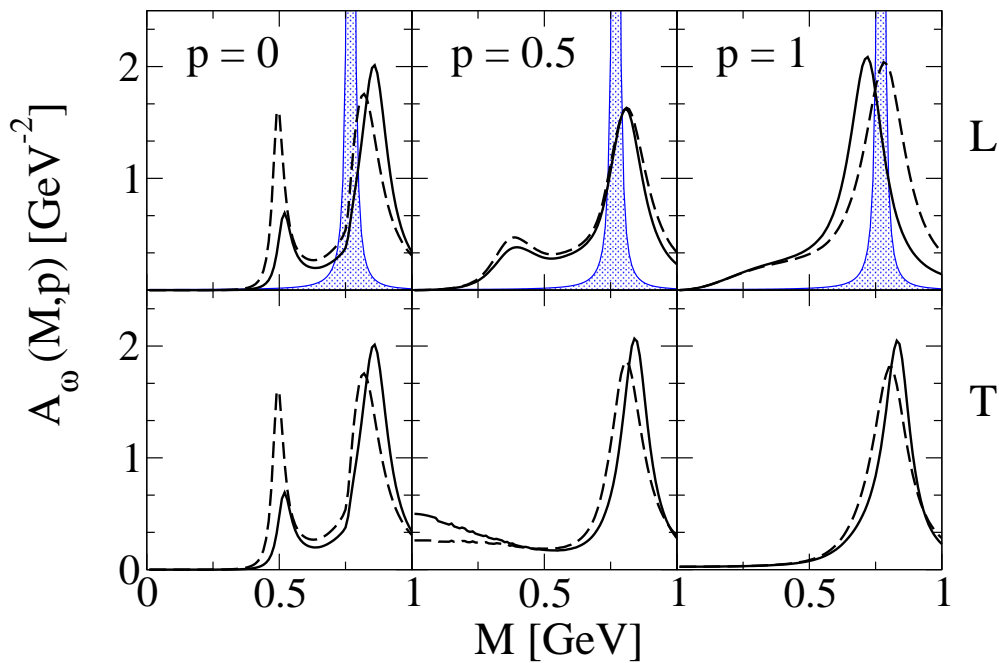}
\caption{(Color online) Same as Fig. \ref{rhomesspecfun}, but for $\omega$ spectral functions.}
\label{omegamesspecfun}
\end{figure}

For the $\omega$ meson, we observe a behavior that is principally similar to 
that of the $\rho$ meson (see Fig.~\ref{omegamesspecfun}). Transverse and 
longitudinal spectral functions are 
similar. In both cases, the $\omega$ pole mass is slightly shifted upward, 
and the $\omega$ is substantially broadened around its quasiparticle pole. 
At $\rho_0$ we obtain an in-medium $\omega$ width of $300$ MeV. 

As in the case of the $\rho$, the coupling to low lying resonances leads 
to the appearance of a first peak  in the spectral function which 
lies around $0.5$-$0.55$ GeV. With increasing momentum, this peak is washed out 
and disappears finally. However, in the case of the  $\omega$, the influence 
of nonresonant contributions is found to be much stronger than for the $\rho$. 
The nonresonant contributions tend to increase 
the repulsive mass shift of the $\omega$ pole, and they strongly 
suppress the first peak in the spectral function. 

This first branch in the spectral distribution is 
mainly generated by the $N^*(1535)$ resonance. As discussed in detail
in Refs. \cite{krivo02,omega}, within the NRD+eVMD model  a strong
$N^*(1535)N\omega$ coupling is implied by the available electroproduction and 
photoproduction data. However, the $N\omega$ decay of this 
resonance has not been measured directly, and therefore input from 
quark model predictions had to be used to fix the entire set of 
eVMD model parameters. Nevertheless, within such a procedure, 
a strong $N^*(1535)N\omega$ coupling seems practically unavoidable. 
In $pp\rightarrow pp\omega$ production, the large $N^*(1535)N\omega$ 
decay mode leads to substantial contributions in a kinematic 
regime where the $\omega$ is far off-shell, i.e., at small invariant 
masses. This is reflected in an enhancement in the cross 
section around threshold \cite{omega}. Existing data 
\cite{hibou99,COSYTOF,Barsov:2006sc}, 
however, do not rule out such a behavior. A closer inspection of the 
experimentally observed background contributions may provide 
important experimental information concerning this question.

\begin{figure}[!htb]
\centering
\includegraphics[width=.6\textwidth]{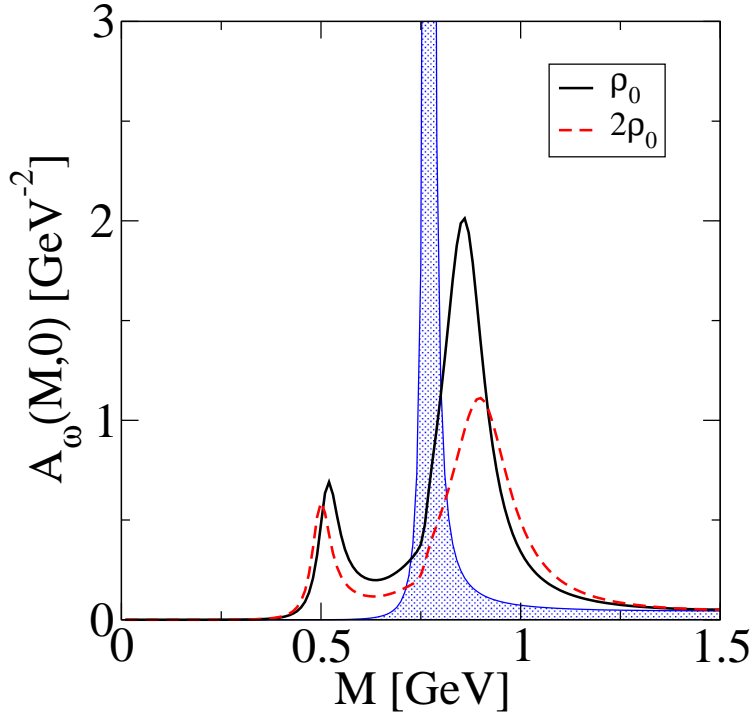}
\caption{(Color online) Same as Fig. \ref{rhomesspecfun1}, but for the unpolarized $\omega$-meson spectral function.}
\label{omegamesspecfun1}
\end{figure}

The nuclear matter density dependence of  the  $\omega$-meson 
spectral function is shown in Fig.~\ref{omegamesspecfun1}. Again, the figure shows the 
unpolarized spectral function at rest at $\rho_0$ and at $2\rho_0$ 
nuclear density. As for the $\rho$ meson, we observe a shift 
of the second peak which belongs to the original $\omega$ pole toward 
higher masses with an increase in density, while the first peak is slightly 
shifted to lower masses. Moreover, the height of the second  
peak is suppressed by about a factor of 2. 

Comparing this with other works, we should mentioned that in the 
pure resonance model approach of Ref. \cite{Pos01}, no such additional 
peak was observed. The $\omega$-meson spectral functions 
obtained within the coupled-channel approach of Ref. \cite{Lutz:2001mi} and 
 within the coupled-channel $K$ matrix of Ref. 
\cite{muehlich06} have qualitative similarity with those from the 
present approach. All approaches come practically to the same 
conclusions: an upward mass shift, a broadening of the $\omega$, 
and the appearance of an additional branch  in the 
$\omega$ spectral function. This branch appears at the same position 
and is in both cases generated by the $N^*(1535)$. However, in all 
approaches, the $\omega$ survives as a quasiparticle, at least at moderate 
densities up to $\rho_0$; i.e., there the spectral function is still 
dominated by the main branch corresponding to the original $\omega$ pole. 
The predictions 
for the density dependence of the spectral function are similar on a 
qualitative level; i.e., when going from one to two times nuclear density, 
the suppression of the branch corresponding to the $\omega$ pole is 
of similar size. 

However, on a quantitative level, the models come to different
conclusions. While the broadening of the $\omega$ is similar in 
Refs. \cite{Lutz:2001mi} and  \cite{muehlich06}, the mass shift is 
much larger in Ref. \cite{Lutz:2001mi} ($\Delta m_\omega\sim 46$ MeV at $\rho_0$) 
than in Ref. \cite{muehlich06} ($\Delta m_\omega\sim 10$ MeV at $\rho_0$). 
In the present case, the in-medium 
modifications of the $\omega$ meson are even more pronounced than in Refs. 
\cite{Lutz:2001mi,muehlich06}; i.e., the broadening and the upward mass shifts are 
larger ($\Delta m_\omega\sim 75$ MeV at $\rho_0$).

A comparison with predictions from QCD sum rules \cite{Zschocke:2002mp,Thomas:2005dc} 
turns out to be difficult because the $\omega$ properties depend strongly 
on higher order condensates. Sum rules leave space for upward and downward 
mass shifts, and the parameters related to the higher order terms in the 
operator product expansion have to finally be fixed from experiments \cite{Thomas:2005dc}. 
Moreover, these approaches assume that the $\omega$ maintains 
its quasiparticle properties. However, due to the distinct 
two-peak structure of the present spectral distributions, it is not 
possible to assign a common mass shift to an $\omega$ quasiparticle pole. 

\subsection{In-medium resonances: Role of self-consistency \label{inmedium_res}}

As the next step, we took into account the changes induced by the in-medium vector mesons 
on the total width of the nucleon resonances. This leads to a 
self-consistent determination of the self-energies of the vector mesons in nuclear matter.

The results shown in the previous section correspond to the first iteration, 
if considered in the context of a self-consistent calculation.
In the second iteration, the in-medium widths of the nucleon resonances 
$\Gamma_R^{*}$ are determined by insertion of the in-medium spectral 
functions of the vector mesons resulting from the first iteration.
Because the latter depend on the momentum of vector meson with respect to 
nuclear medium ${\bf p}$, 
the in-medium widths of the nucleon resonances $\Gamma_R^{*}$ will  
depend on the resonance momentum $|{\bf p}_R|$, that is,
\beq
\Gamma_R^{*}(s,|{\bf p}_R|) = 
\Gamma_R^{\mathrm{tot}}(s)+\sum_{V}\int\Gamma_{RNV}(s,M)\Delta{\cal A}_V 
(M,|{\bf p}| )dM^2\frac{d\Omega}{4\pi}~,
\label{selfconsist}
\eeq
where $\Delta{\cal A}_V$ refers to the modification of vector meson spectral 
functions with respect to the
vacuum ones. $|{\bf p}|$, among other things, depends on $|{\bf p}_R|$ and on the 
orientation of the decay products momenta with respect
to the direction of the resonance momentum.

Doing so, nucleon-resonance scattering terms leading to the broadening 
of the resonances are produced~\cite{leupold04}. 
The vector meson self-energies are then calculated from Eqs. (\ref{sigma})--(\ref{ampl1}) 
using $\Gamma_R^{*}(s,|{\bf p}_R|)$ instead of $\Gamma_R^{\mathrm{tot}}(s)$.
The procedure is repeated until convergence. 
We find that the convergence is obtained after the third iteration.

As a side result of our self-consistent calculation, we find that the widths 
of the nucleon resonances are enhanced in medium because the vector 
meson spectral functions show a significative spectral strength at small invariant masses. 
A similar outcome emerged from the analysis performed in Ref.~\cite{postrho1}.

\begin{figure}[!htb]
\centering
\includegraphics[width=.9\textwidth]{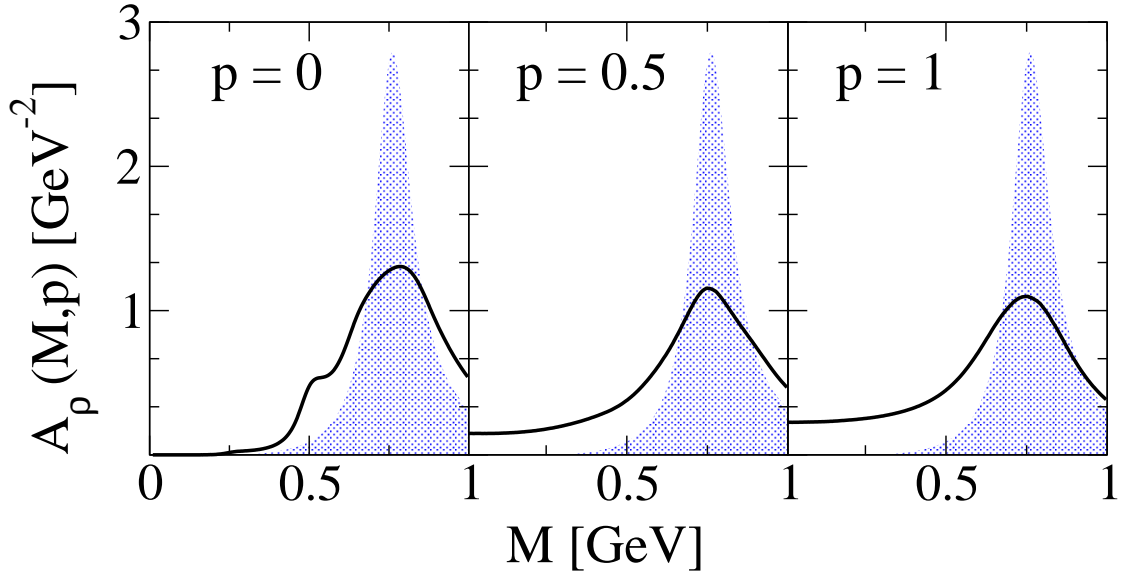}
\caption{(Color online) Unpolarized spectral functions of the $\rho$ meson
in nuclear matter at saturation density for various momenta $p$ (in
GeV). The broadening of the nucleon resonance widths induced by the 
in-medium spectral properties of the vector mesons 
is taken into account and a self-consistent calculation is performed.
The shaded area shows the vacuum
spectral function.}
\label{rhospf.SC}
\end{figure}

\begin{figure}[!htb]
\centering
\includegraphics[width=.9\textwidth]{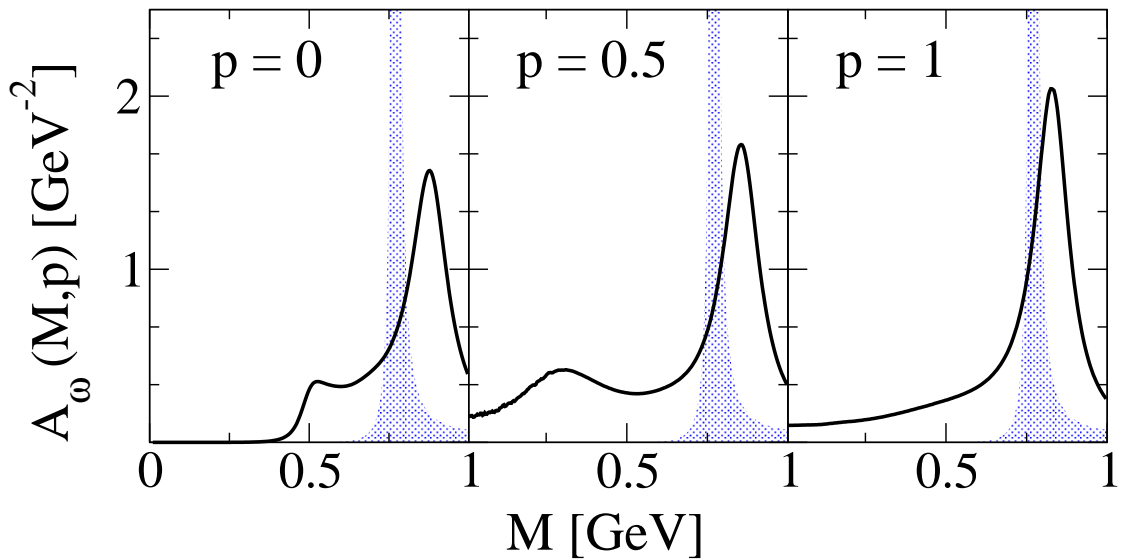}
\caption{(Color online) Same as Fig. \ref{rhospf.SC}, but for the $\omega$ meson.}
\label{omespf.SC}
\end{figure}

The resulting unpolarized vector meson spectral 
functions are shown in Figs.~\ref{rhospf.SC} and \ref{omespf.SC} for 
the $\rho$ and $\omega$ mesons, respectively. They 
refer to saturation density. We observe that the self-consistent 
calculation leads predominantly to a reduction of the lower mass peak. 
This result qualitatively agrees with the findings 
of Ref.~\cite{leupold04}, which investigated the role of a self-consistent iteration scheme 
on the $\rho$-meson spectral 
function.

\subsection{Experimental situation}

Experimental constraints on the in-medium $\omega$ spectral function can 
presently be derived from the CB-TAPS ($\gamma+ A$) experiments \cite{cbelsa}, 
the $p+A$ measurements at KEK \cite{KEK,KEK2}, and the heavy-ion dilepton 
experiments. The dilepton measurements of the DLS Collaboration in C+C and 
Ca+Ca at 1$A$ GeV \cite{Por97} suffer from too low mass resolution in the 
vicinity of the $\omega$ peak in order 
to make precise statements on the $\omega$ in-medium width. However, there 
is no doubt that the explanation of the DLS data requires a substantial 
broadening of the  $\omega$ spectral function. The analysis of Ref. \cite{She03} 
showed that the DLS data are compatible with a rather large $\omega$ width; 
i.e., $\Gamma_\omega^{\mathrm{tot}}\sim 150$--$300$ MeV. 
The first data from HADES 
\cite{Had06} will be analyzed in the next section. 

As discussed in Ref. \cite{Elena}, the interpretation of the $p$+C and $p$+Cu 
KEK dilepton data \cite{KEK,KEK2} suffers from the high initial proton kinetic energy of 
12 GeV. This means that vector mesons are produced with high momenta ($p>1$ GeV) and 
in particular the $\omega$ decays at low nuclear densities or even outside 
the target nucleus. Nevertheless, the KEK study observed a substantial 
difference of the dilepton spectrum with respect to the standard sources 
below the $\rho/\omega$ peak. On a qualitative level, these data support 
a picture as predicted by the resonance model, i.e., a substantial 
shift of spectral strength to smaller masses.

However, the resonance model and coupled-channel predictions for the $\omega$ 
spectral function contradict, at least partially, the results of a recent 
photoproduction experiment carried out by the CBELSA/TAPS Collaboration 
\cite{cbelsa}.  The experiment indicates a downward mass shift of about 
$\Delta m_\omega\sim - 80$ MeV at $\rho_0$, whereas the coupled-channel 
\cite{Lutz:2001mi,muehlich06} and resonance model calculations predict 
a more or less pronounced repulsive shift ranging from +10 to +80 MeV. The 
collisional broadening  extracted by CBELSA/TAPS is moderate; i.e., 
$\Delta \Gamma_\omega\sim 50$ MeV. Since this experiment has been carried out 
at beam energies of 0.64--2.53 GeV, a similar argument as for the KEK experiment applies, 
at least for the high energies, namely, that one has to
carefully  account for the energy dependence of the $\omega$ 
spectral function and to follow the paths of the $\omega$ decays. 

Recent $\gamma+A$ measurements from the CLAS Collaboration \cite{CLAS06} at the 
Thomas Jefferson National Accelerator Facility, carried out at photon energies 
$E_\gamma =0.6$--$3.8$ GeV  on 
light (C) and heavier (Fe) targets, find no signatures for a vector meson 
mass shift for the $\rho$ and $\omega$ mesons but indicate a collisional 
broadening of the $\rho$ by $50$--$70$ MeV.   

In terms of a simple BR scaling interpretation $m_V^* = m_V(1 - \alpha\rho_B/\rho_0)$ 
which should hold in common for both the $\rho$ and the $\omega$ meson, 
the best fits to the various experiments yield at present a divergent picture: 
$|\alpha| \sim 0.13$  (CBELSA/TAPS \cite{cbelsa}), 
$|\alpha| \sim 0.092\pm 0.002$ (KEK-PS E325 \cite{KEK2}), and 
$|\alpha| \sim 0.02\pm 0.02$ (CLAS \cite{CLAS06}). 

This apparent contradiction implies already that the mass shift scenario 
\`a la Brown-Rho  is too simple, and, consistent with the NA60 heavy-ion data \cite{na60}, 
the vector mesons develop more complex spectral properties.

\subsection{Realization within the transport approach}

A first attempt to introduce in-medium spectral functions of the vector mesons 
in a transport description for intermediate energy heavy-ion collisions 
was performed in Ref.~\cite{BCRW98}. The in-medium dilepton rate was thereby 
expressed in terms of the $\rho$ meson in-medium spectral function. 
The proportionality was achieved by including the medium effects 
at the level of production channels.
However, to avoid double counting, this required 
switching off the explicit  
$\rho$ meson production channels included in the self-energy calculations, 
which have been implicitly accounted for in terms 
of the in-medium spectral function. 
In particular, the decays of the $\rho$ mesons 
produced in baryon-baryon collisions and meson-baryon 
interactions were not included explicitly.

In the language of a resonance model, this would mean, e.g., 
that since $V+N\rightarrow R$ is a 
(dominant!) contribution to the self-energy, 
the $R\rightarrow NV\rightarrow Ne^+e^-$ decay should 
not be included explicitly, if one would operate as in Ref.~\cite{BCRW98}. 
However, nucleon resonances are important dynamical degrees of 
freedom of a transport approach.
In particular, in heavy-ion collisions at intermediate energies, the 
medium is dominated by nucleons and nucleon resonances, and to 
neglect an explicit 
(dynamical) treatment of the latter in the 
determination of vector meson production to 
obtain a direct proportionality to the vector meson spectral function 
is a questionable procedure.
Moreover, it was already pointed out in Ref~\cite{rapp00}  that such a  
treatment ~\cite{BCRW98} might not be realistic.

Here we adopt an alternative approach to extract information 
on the in-medium vector meson spectral functions from dilepton 
emission in heavy-ion collisions.
The idea is to restrict propagation and mutual interactions to  
the dynamical degrees of freedom  within the transport approach.  
Vector mesons and their interactions are treated as  
perturbative degrees of freedom within the transport description. This gives us 
the possibility of including the corresponding vector meson 
in-medium modifications on a microscopic level. The philosophy behind 
this approach is similar to that of the approach pursued in 
Ref. \cite{rapp00,vanHees:2006ng}, which derived local dilepton emission rates 
from the decay rates of 
the corresponding  sources within the 
framework of an expanding fireball model.

To be more precise, within the NRD model the vector meson production 
channels are 
nucleon resonances which are treated as explicit, i.e., as dynamical 
degrees of freedom 
in the transport code up their decay. Dilepton emission takes place 
via resonance Dalitz decays (eVMD)  where the 
vector mesons enter as virtual particles the $RN\gamma^*$ vertices.  
In the medium, the vector mesons entering into the vertex form factors are modified 
by the vector meson self-energy, determined for the conditions, i.e., at the 
density where the resonance decay takes place. Thus the presence of the medium 
changes the branching ratios for the nucleon resonance Dalitz 
decays. This modification, however, is 
not directly proportional to the vector meson spectral function 
but rather to the in-medium form factors.
In this approach, e.g., vector meson absorption processes are taken into account 
microscopically, although not dynamically, in terms of the 
imaginary part of the vector meson self-energy, which determines the 
corresponding collisional broadening.

By this procedure, one avoids the complicated and yet not fully resolved 
problem of a consistent off-shell propagation of the vector mesons within 
semiclassical transport models \cite{BCRW98,Barz:2006sh}. 
A drawback is certainly that one loses in this picture 
information on the dynamical propagation of the vector mesons. Consequently,   
their in-medium properties are determined by the conditions at the decay 
points of the nucleon resonances. Since the vector mesons appear only as 
intermediate states in the resonance decay rates, it is, on the other 
hand, much easier to include spectral functions and to keep quantum 
effects which are lost in semiclassical approaches, even when off-shell 
effects are taken into account \cite{Leupold:2000ma,cassing00}.

The in-medium spectral properties discussed in the next subsection 
are determined in a local density approximation (LDA). In RQMD, the local 
baryon density $\rho_B$ is determined by the summation over Gaussian wave 
packets of all nucleons and resonances (it should not be mixed up with 
the interaction density used in RQMD for the determination of the intranuclear 
forces \cite{ai91}). Thus the determination of the baryon density does 
not require local equilibrium as in hydrodynamics, 
but dependences on particular 
nonequilibrium effects such as phase-space anisotropies \cite{Fuchs:2002uw} or 
memory effects \cite{Schenke:2006uh} are neglected. As usually done at 
intermediate energies, an explicit temperature dependence of spectral properties is 
neglected as well. It is, however, possible to extract local temperatures 
from transport simulations, either by fitting hot Fermi distributions to 
local momentum space configurations \cite{Fuchs:1997we} or by performing 
thermal model fits to local hadron abundances and spectra \cite{Bravina:2008ra}. 
Although both procedures are connected with an extremely high numerical 
effort, it is thus possible to include density- {\it and} 
temperature-dependent spectral functions into transport simulations. 
Since temperature 
effects have been found to dominate  at SPS energies 
\cite{vanHees:2006ng}, such an extension will be necessary for an
 application of the present model to, e.g., NA60 data.

\subsection{In-medium dilepton emission rates}

Because of the $P$ invariance of the electromagnetic interaction, resonances 
with arbitrary spin have only 
three independent helicity amplitudes in the $\gamma ^{*}N\rightarrow R$ 
transitions. This means that there are three independent scalar functions 
to fix the vertices. The three scalar functions arising 
from the decomposition of the $\gamma ^{*}N\rightarrow R$ vertex 
over the Lorentz vectors and the Dirac matrices are functions of the mass 
squared $M^2$ of the virtual photon and are called covariant form factors. 
In the eVMD model, each of these covariant form factors is 
expressed in a gauge invariant way (see Appendix) as a 
linear superposition of the 
contributions from the intermediate vector mesons of the $\rho$ and $\omega$ family. In 
contrast to the naive VMD, in which only the  $\rho$ and $\omega$ ground states 
are taken into account, eVMD includes radial excitations 
$\rho (1450)$, $\rho (1700)$, etc., 
 which interfere with the ground-state $\rho$ mesons in radiative processes.
The corresponding transition form factors are given by \cite{krivo02}

\begin{eqnarray}
F_{k}^{(\pm)}(M^2)=\sum_{i}\,{\mathcal M}_{ki}^{(\pm)}~,
\label{formfac}
\end{eqnarray}
where $k=1,\ldots,3$ stands for each of the form factors, $(\pm)$ 
denotes states of normal and abnormal parity, respectively, and 
the sum is over the intermediate mesons. The $\Delta $  resonance form 
factors have contributions from only the $\rho$-meson family, 
whereas the nucleon resonances receive contributions 
from the $\rho $ and $\omega $ 
mesons. For a resonance of spin $J=l+1/2$, the total number of vector 
mesons is $l+3$.
The amplitude
\begin{eqnarray}
{\mathcal M}_{k,i}^{(\pm)}=h_{ki}^{(\pm)}\,\frac{m_i^2}{m_i^2-im_i\Gamma_i-M^2}
\label{amplM}
\end{eqnarray}
represents the contribution of the $i$th vector meson to the 
form factor of type $k$.
The residues $h_{ki}^{(\pm)}$ contain the free parameters of the model. 
They are constrained by the requirement
that the asymptotic expression of the form factors is consistent 
with the quark counting rules~\cite{Bro73}. For each form factor, the
quark counting rules reduce the number of free parameters from ${l+3}$ 
to $2$ for $k=1$ and to 
$1$ for $k=2,3$. The remaining parameters are fixed
by fitting the available photoproduction and electroproduction data and 
using results
of the multichannel partial-wave analysis of the $\pi N$ scattering. Where
experimental data are not available, predictions of the nonrelativistic
quark models are used as an input.

The $\Gamma (R\rightarrow N\gamma ^{*})$ decay width can be written in 
terms of three transition form factors (magnetic, electric, and Coulomb) 
for a resonance with spin $J>1/2$ and two for $J=1/2$. 
The matrix elements connecting the former with the covariant form factors 
are 
explicitly listed in Ref.~\cite{krivo02}.

In this representation, the insertion of the in-medium properties of the 
$\omega$ and $\rho$
 vector mesons is straightforward. In the medium, the transition amplitudes 
${\mathcal M}_{k, i}^{(\pm)}$ ($i=\rho$, $\omega$, $\dots$) are 
directly modified by the in-medium self-energy  and read
\begin{eqnarray}
{\mathcal M}_{k, i=V}^{(\pm)}=h_{kV}^{(\pm)}\,
\frac{m_V^2+\Re\Sigma_V^{\mathrm{tot}}}{m_V^2+\Re\Sigma_V^{\mathrm{tot}}+i\Im\Sigma_V^{\mathrm{tot}}-M^2}~.
\label{inmedampl}
\end{eqnarray}
We include the self-energy contributions for the ground-state $\rho$ and $\omega$ mesons 
in the transition. For the excited states 
$\rho^\prime,\rho^{\prime\prime},\dots$, the 
self-energies are unknown, and thus we keep for these states their vacuum properties.

As in Ref. \cite{Cozma06}, we also consider scenarios in which the  
self-energy is based on different model assumptions, namely, a simple 
Brown-Rho (BR) or Hatsuda-Lee 
scaling of the vector meson masses \cite{Bro91,Hat92} and a collisional 
broadening 
of the  vector meson widths. In the latter case, the self-energies are given by 
\beqa
\Im\Sigma_V^{\mathrm{tot}} &=& - m_V \left(\Gamma_V^{(0)}(M)+\Gamma_V^{\mathrm{coll}}(\rho_B,M)\right)~,
\nonumber\\   
\Re\Sigma_V^{\mathrm{tot}} &=& 0  ~.
\label{brcoll}
\eeqa
In this context, we want to stress that in Eqs. (\ref{brcoll}), the energy dependence due to the 
two- or three-pion decay of the vector meson is kept in the vacuum 
contribution to the total width, while the collisional broadening due to
 the interaction 
with the surrounding nucleons is absorbed into a density- and energy-dependent 
part. The issue of the energy dependence of the 
collisional width will be discussed in detail in the next section.
The BR scaling 
is introduced through the
replacement $m_V \rightarrow m^*_V=m_V (1-\alpha\frac{\rho_B}{\rho_0})$, 
as done, e.g., in Ref.~\cite{Li:1996mi}. In particular, in this case, one has
\beqa
\Re\Sigma_V^{\mathrm{tot}} &=& \left(m_V -\alpha\frac{\rho_B}{\rho_0}\right)^2-m_V^2~.
\eeqa
 As usual, the mass shift entering into the real part can be adjusted by 
the parameter $\alpha$. As in the case of full spectral functions, 
the self-energy 
components enter into the amplitudes (\ref{inmedampl}).
In this context, it is important to note that the modification of 
the amplitudes 
(\ref{inmedampl}) leads to a {\it coherent} summation of the  
$\rho$ and $\omega$ spectral 
functions in the transition form factors (\ref{formfac}). Doing so, 
this approach 
goes beyond the standard--even off-shell--transport approach 
where spectral properties 
are treated at the level of cross sections 
\cite{BCRW98,Leupold:2000ma,Barz:2006sh}. 
The latter always leads to an incoherent summation of the 
contributions from different 
hadrons.

The self-energy appearing in Eq.~(\ref{inmedampl}) is a 
function $\Sigma_V^{\mathrm{tot}}(M,|{\mathbf p}|,\rho_B)$ 
of the vector meson running mass, the modulus of its three-momentum 
in the nuclear matter rest frame, and the local 
density of the surrounding matter.
In the rest frame $L^{*}$ of a resonance $R$ with mass $\mu$, decaying into a 
nucleon and a vector meson of mass $M$, the modulus $|{\mathbf p^{*}}|$ of the 
momentum of the meson is fixed by energy conservation. If ${\mathbf p}_R$ is the 
momentum of the resonance $R$ in the c.m. frame $L$ of the colliding nuclei and 
${\mathbf v}_R={\mathbf p}_R/\sqrt{{\mathbf p}_R^2+\mu^2}$ is its velocity, the 
vector meson momentum in $L$ 
is given by the Lorentz transformation
\begin{equation}
|{\mathbf p}|^2=(\gamma_R|{\mathbf v}_R|E^*+\gamma_R p^{*}_L)^2+p^{*2}_T~,
\end{equation}
where
\begin{eqnarray}
p^{*}_L&=&|{\mathbf p^{*}}|\cos\theta~, \\
p^{*}_T&=&|{\mathbf p^{*}}|\sin\theta~,
\end{eqnarray}
with $\theta$ being the polar angle of the meson in $L^{*}$ if one chooses the 
$z$ axis 
of this frame pointing in the direction of ${\mathbf v}_R$.
Since $|{\mathbf p^{*}}|$ is fixed, in terms of the $L$ 
frame variables, one has $\Sigma_V^{\mathrm{tot}}=\Sigma_V^{\mathrm{tot}}(M,\cos\theta,\rho_B)$, 
and the decay amplitude averaged over the angles reads
\begin{equation}
\Gamma (R\rightarrow N\gamma ^{*})(\mu,M,\rho_B)=\int_{-1}^{+1}\frac{d\cos(\theta)}{2}\Gamma (R\rightarrow N\gamma ^{*})(\mu,M,\cos\theta,\rho_B)~.
\label{averampl}
\end{equation}
Equation (\ref{averampl}) is implementable in the framework of the T\"ubingen 
RQMD transport code. 
The RQMD code \cite{Uma98,She03,fuchs05} has been 
extended to include all nuclear resonances with masses below 2 GeV, 
in total 11 $N^*$ and 10 $\Delta$ resonances.
A full list with the corresponding masses and decay widths to various 
channels can be found in Tables III and IV of Ref.~\cite{She03}.
For each resonance, RQMD provides the values of the three-momentum components 
(necessary to perform the Lorentz boost), the mass (distributed over a Breit-Wigner), and the local density of the surrounding matter at the decay point. 

Since vector mesons play in the eVMD model the role of intermediate 
virtual particles, their off-shellness is fully taken into account in a consistent manner.

The model can be applied to dilepton production in heavy-ion reactions.
In the energy range of a few $A$ GeV, one can identify three main classes of 
processes that lead to dilepton emission: nucleon-nucleon bremsstrahlung, 
decay of light unflavored mesons, and decay of nucleon and $\Delta$ resonances. 
Dilepton production through the bremsstrahlung mechanism 
has been studied in detail in Ref.~\cite{Shyam:2003cn}. For the energy
range of interest in this work, bremsstrahlung contributes in a significant way only 
at small invariant masses  to the dilepton spectrum. 
By far, the dominant contributions result from diagrams that 
involve the excitation of an intermediate $\Delta$ resonance. Within 
the present framework, the inclusion of such contributions would, however, 
lead to a double counting and therefore we omit explicit bremsstrahlung
contributions. Recently, the quantitative importance of 
bremsstrahlung contributions has again been discussed in Ref.~\cite{Kaptari:2005qz}, 
however, with results contradictory to those in Ref. \cite{Shyam:2003cn}.

At incident energies of a few $A$ GeV, the cross sections for meson 
$\mathcal{M}=\eta,\eta',\rho,\omega,\phi$ production are small, and 
these mesons do not play an important role in the dynamics of heavy-ion
collisions. Their production can thus be treated perturbatively, in contrast 
to the case of the pion. The
decay to a dilepton pair takes place through the emission of a virtual photon. 
The differential branching
ratios for the decay of a meson to a final state $Xe^+e^-$ can be written 
\begin{eqnarray}
dB(\mu,M)^{{\mathcal M},\pi\rightarrow e^+e^-X}=\frac{d\Gamma(\mu,M)^{{\mathcal M},\pi\rightarrow e^+e^-X}}
{\Gamma_{\mathrm{tot}}^{{\mathcal M},\pi}(\mu)}\:,
\end{eqnarray}
with $\mu$ the meson mass and $M$ the dilepton mass. Three types of such 
decays have been considered: direct
decays ${\mathcal M}\rightarrow e^{+}e^{-}$, Dalitz decays 
${\mathcal M}\rightarrow\gamma e^{+}e^{-}$ and ${\mathcal M}\rightarrow\pi(\eta) e^{+}e^{-}$, and four-body decays
${\mathcal M}\rightarrow\pi\pi e^{+}e^{-}$. A comprehensive study of the decay of
light mesons to a dilepton pair has been performed in Ref.~\cite{Fae00}. 
Assuming a NRD model for the production of $\rho$ and $\omega$ mesons, 
the remaining decay channels that are most important quantitatively for 
heavy-ion collisions at 1$A$ and 2$A$ GeV are $\pi^0\rightarrow \gamma e^+e^-$ and
$\eta\rightarrow \gamma e^+e^-$.

In terms of the branching 
ratios for the Dalitz decays of the
baryon resonances, the cross section for $e^+ e^-$ production from 
the initial state $X^\prime$ together with the final state 
$N X$ can be written as 
\begin{equation}
\frac{d\sigma (s,M)^{X^{\prime} \rightarrow N X e^{+}e^{-}}}{dM^{2}}
=\sum_{R}\int_{(m_{N}+M)^{2}}^{(\sqrt{s}-m_{X})^{2}}d\mu ^{2}
\frac{d\sigma (s,\mu )^{X^\prime \rightarrow R X}}{d\mu ^{2}}\sum_{V} 
\frac{dB(\mu,M,\rho_B)^{R\rightarrow VN\rightarrow N e^{+}e^{-}}}{dM^{2}}~~.
\label{1}
\end{equation}
Here, $\mu $ is the running mass of the baryon resonance $R$ with the cross
section $d\sigma (s,\mu )^{X^\prime\rightarrow X R}$, 
$dB(\mu ,M,\rho_B)^{R\rightarrow VN\rightarrow N e^{+}e^{-}}$ is the 
differential branching ratio for the
Dalitz decay $R\rightarrow N e^{+}e^{-}$ through the vector meson $V$. Thus 
Eq.~(\ref{1}) describes baryon- and pion-induced dilepton production; i.e., 
the initial state can be given by two baryons 
$X^\prime = NN,~ NR,~ R^\prime R$ or 
it runs through pion absorption $X^\prime = \pi N$. In the resonance model, 
both processes are treated on the same footing by the decay of intermediate 
resonances. Medium modifications enter the branching ratio $dB(\mu
,M,\rho_B)^{R\rightarrow VN\rightarrow N e^{+}e^{-}}$ 
by affecting the Dalitz decay width $d\Gamma(\mu
,M,\rho_B)^{R\rightarrow VN\rightarrow N e^{+}e^{-}}/dM^2$.
Once the $\Gamma (R\rightarrow N\gamma ^{*})$ is calculated within the eVMD, the
factorization prescription~\cite{krivo02} can be
used to find the dilepton decay rate
\begin{equation}
d\Gamma (R\rightarrow Ne^{+}e^{-})=\Gamma (R\rightarrow N\gamma
^{*})M\Gamma (\gamma ^{*}\rightarrow e^{+}e^{-})\frac{dM^{2}}{\pi M^{4}},
\label{OK!}
\end{equation}
where 
\begin{equation}
M\Gamma (\gamma ^{*}\rightarrow e^{+}e^{-})=\frac{\alpha }{3}%
(M^{2}+2m_{e}^{2})\sqrt{1-\frac{4m_{e}^{2}}{M^{2}}}  \label{OK!!}
\end{equation}
is the decay width of a virtual photon $\gamma ^{*}$ into the dilepton
pair with the invariant mass $M$. 

As discussed in Ref.~\cite{leupold04}, the excitation of particle-hole 
pairs in the meson spectral function generates resonance-nucleon scattering 
terms in the resonance self-energy and thus the in-medium broadening of the resonance. 
We have mentioned that nucleon resonances are dynamically treated in the RQMD model, and 
resonance-nucleon scattering is explicitly performed. Thus, 
the in-medium broadening of nucleon 
resonances is taken into account in the transport approach 
\emph{dynamically}.
No in-medium spectral functions of the 
vector mesons are therefore included in the total width
$\Gamma_{\mathrm{tot}}^{R}(\mu)$ entering in the branching ratio in 
Eq.~(\ref{1}). 

The real part of the resonance self-energy $\Re \Sigma_{\mathrm{tot}}^{R}$ 
is included in a phenomenological way through a 
mean field in which the resonances are propagated. It is therefore 
assumed that the nucleon resonances feel the same potential as nucleons, which 
is a standard approximation in present transport models and should be improved 
in the future. For example, the $RNV$ vertex gives rise to additional Fock 
contributions which could be included in future work.

An observable tightly connected to a correct treatment of  the resonance dynamics 
in heavy-ion collision transport calculations is provided by the pion multiplicity. 
 For the mass system under 
consideration, pion multiplicities are reasonably well reproduced
 by the present description. For example, inclusive $\pi^+$ cross sections in 
C+C reactions measured by the KaoS Collaboration 
\cite{Stu01} can be reproduced by 
the present description within error bars.
This gives, at least on a global level, 
manifest credit to our treatment.

For the $\eta $, we include $\eta$ absorption from the dominating 
channel $\eta+N \rightarrow N^*(1535)$ explicitly. Since 
chiral perturbation theory predicts practically no 
modifications of the in-medium $\eta $ mass \cite{oset02}, we do 
not include a possible $\eta$  mass shift.

\begin{figure}[!htb]
\centering
\includegraphics[width=.5\textwidth]{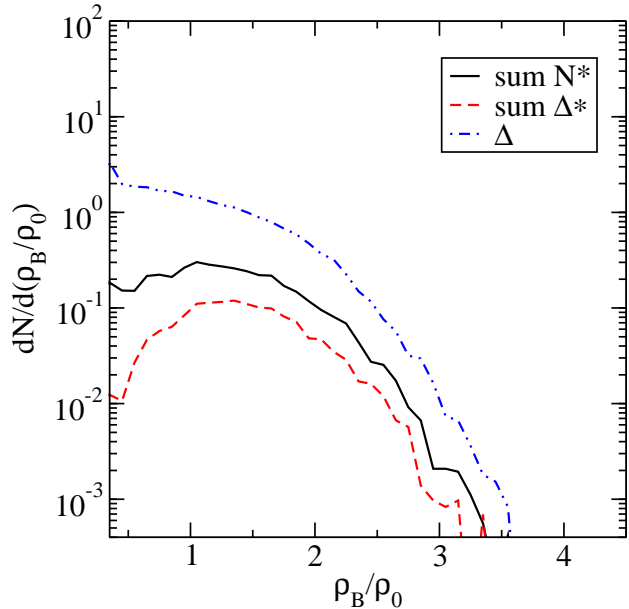}
\caption{(Color online) Impact parameter weighted 
density distribution of the nucleon resonances 
taken into account in the calculations at their decay point. 
The distribution of the $N^*$, $\Delta^*$, and $\Delta(1232)$ 
resonances are separately shown.}
\label{dndrho_res}
\end{figure}

To give an impression of the density range relevant for
dilepton production in the C+C system, Fig.~\ref{dndrho_res} 
shows the density distribution $dN/d\rho_B$ where the nucleon resonance 
decays into dilepton channels take place. Note that Fig. \ref{dndrho_res} 
refers to minimal bias conditions. Triggering on central reactions 
and/or increasing the system size will
help to better explore the high density range. To have a
separate look at $\rho$ and $\omega$ production, we distinguish
between  $N^*$, $\Delta^*$, and $\Delta(1232)$ 
resonances. It can be seen that the highest mass resonances, i.e., 
$N^*$ and $\Delta^*$, decay at supranormal densities, while 
a large fraction of the  $\Delta (1232)$ decays 
take place at lower densities, between $0.5\rho_0$ and $1\rho_0$.
In all cases, however, maximal densities up to $3\rho_0$--$4\rho_0$ are reached.
Thus, already the small C+C system probes the spectral properties of 
intermediate vector mesons from  $N^*$ and $\Delta^*$ decays at 
supranormal densities.

\section{Dilepton production}

In this section, we provide theoretical calculations of the dilepton
emission in heavy-ion collisions at intermediate energy. 
In particular, we address the reaction C+C at 2$A$ GeV for 
which experimental data have been 
already released by the HADES Collaboration. The main purpose is to 
compare calculations that include in-medium effects in a more 
traditional way, i.e.,  via Brown-Rho scaling of the vector 
masses and empirical collisional broadening of the decay width, 
with results obtained using $\rho$ and $\omega$ mesons described 
by the in-medium spectral functions of the previous section. 
New HADES data \cite{Agakishiev:2007ts} will be analyzed elsewhere.

In the transport calculation, the reaction 
has been treated as minimal bias collisions with respective
 maximal 
impact parameter $b_{\mathrm{max}}= 6.0$ fm. For the nuclear 
mean field, 
a soft momentum-dependent Skyrme force ($K=200$ MeV) is used \cite{ai91}
which provides also a good description of the subthreshold $K^+$ 
production in the considered energy range \cite{fuchs01}.  
To perform the comparison with the HADES data, 
dilepton events originated from the different considered sources
 were generated in the phase space. After 
smearing over the experimental momentum resolution,
the acceptance filter function provided by the HADES Collaboration 
was applied. Events with opening angle $\theta_{e^+ e^-} \leq 9^{\circ}$ 
were rejected, 
in accordance with the treatment of the experimental data. The 
spectrum was then normalized to the corresponding $\pi^0$ multiplicity.
\

\subsubsection{Vacuum}

\begin{figure}[!htb]
\centering
\includegraphics[width=.45\textwidth]{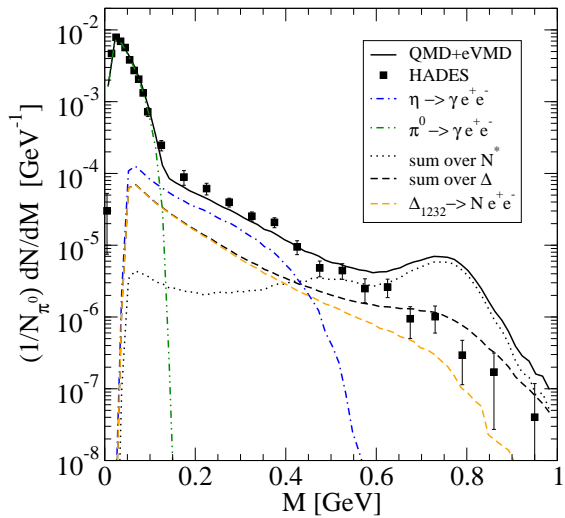}
\caption{(Color online) Dilepton spectrum in C+C reaction at 2.0$A$ GeV as predicted 
by the vacuum NRD+eVMD model and compared with HADES data~\cite{Had06}. The contribution of the different
 types of sources taken into account 
in the calculation is explicitly shown.
}
\label{vacuum}
\end{figure}

We start by addressing the results obtained without any additional 
medium effects concerning the dilepton production.
In Fig.~\ref{vacuum},
 the dilepton spectrum obtained within the vacuum formulation 
of the NRD+eVMD model is compared with the HADES data~\cite{Had06}.
The experimental data 
are slightly underestimated in the mass region $m_\pi\leq M \leq 0.4$ GeV
 and overestimated in the region of the vector meson peak. 
Indeed, the comparison with DLS data had already shown that the 
eVMD model in its pure vacuum formulation fails in describing dilepton 
production in heavy-ion
collisions~\cite{She03}. However, the vacuum calculation is a good reference 
point for isolating, where possible,
 those sources that dominantly contribute to the 
spectrum in a certain invariant mass region.
Once the dominant sources have been individuated, it is interesting to look 
separately at their modifications due to in-medium effects.
For this purpose, we also show separately in Fig.~\ref{vacuum} 
the contributions to the 
spectrum of the decays
of the pseudoscalar $\eta$ and $\pi^0$ mesons and all the 
$N^*$ as well as the $\Delta$ resonances. In addition, the 
$\Delta(1232)\rightarrow N e^+e^-$ decay channel is explicitly shown. 
In what follows, we will investigate the modification of the 
Dalitz decays of the baryon resonances 
due to the introduction of the in-medium 
properties of the $\rho$ and $\omega$ mesons. 
Since we introduce no in-medium modifications of the 
$\pi^0 \rightarrow \gamma e^+e^-$ and $\eta^0 \rightarrow \gamma e^+e^-$ 
channels, the contribution to the dilepton spectrum from the $\pi^0$ and 
$\eta$ Dalitz decay will remain unchanged in the course of our analysis.

The $\eta$ multiplicity $M_{\eta}(4\pi)[10^{-4}]=330$ for the considered 
reaction C+C at 2$A$ GeV is in agreement with experimental data from 
TAPS \cite{Averbeck:1997ma}: $M_{\eta}(4\pi) [10^{-4}]=294 \pm 46 $. 

Before coming to the discussion of medium effects, the vacuum results should 
be examined more closely. As compared to that reported in Ref. \cite{Bratkovskaya:2007jk}, we 
find a higher yield around the $\rho/\omega$ peak region. This enhancement 
arises in the present model first because of a strong coupling of the $\omega$ 
to the $N^* (1535)$ resonance as discussed in detail in Refs. \cite{She03,Cozma06}.
Second, additional enhancement results from the implementation of the quark counting 
rules to the nucleon resonance transition form factors. 
The quark counting rules are known 
to be a well-founded consequence of QCD and, furthermore, are required experimentally 
to match the photon and $\rho$-meson branchings of the nucleon resonances also \cite{resdec}. 

A precise estimate of the $\omega$ contribution is particularly 
important for extracting the 
$\omega$ meson in-medium width: the underestimation of the dilepton yield 
gives rise to the underestimation of the width. 
As we shall see, the in-medium $\omega$ peak is 
strongly suppressed because of the $\omega$ meson broadening.

The low-mass region is critical for understanding the DLS puzzle. In the present 
vacuum calculation,  we obtain a low-mass dilepton yield that is about a 
factor of 2 smaller than that in Ref. \cite{Bratkovskaya:2007jk}. 
Bremsstrahlung cannot explain this deviation, since at 2$A$ GeV it is marginal whatever 
maximal cross sections under debate \cite{Kaptari:2005qz} would have been used. 
At small $M$, the variances in predictions of the present transport models arise from two 
additional sources, namely, the $\eta$ contribution and the $\Delta$ Dalitz decay: For the $\eta$ 
decay which dominates the low-mass dilepton yield 
\cite{She03,Cozma06,Bratkovskaya:2007jk,Thomere:2007cj,Schumacher:2006wc}, we obtain quite standard values. 
The main difference lies therefore in the $\Delta$ Dalitz decay. The present calculation is close 
to that in Ref. \cite{Thomere:2007cj} and about a factor of 5  lower than the calculation 
of Ref. \cite{Bratkovskaya:2007jk}. This point is crucial, since the whole interpretation of 
the low-mass dilepton spectra depends on this fact.

The problems on the  $\Delta(1232)$ Dalitz decay have occurred already at a kinematic level where 
theoretical calculations of several groups surprisingly disagree with each other
pairwise (for a detailed discussion, see Ref. \cite{Krivoruchenko:2001hs}). The dilepton 
decays can be determined from the radiative decays by factorization, which means that 
the $N \gamma^* \mapsto \Delta $ amplitudes have to be determined first. In this context, it should be noted that from all nucleon resonances, the $N \gamma^* \mapsto\Delta (1232)$
transition amplitudes are the best constrained from the experimental point of view 
(see, e.g., Fig. 20 in Ref.~\cite{krivo02}). The $\Delta(1232)$ Dalitz decay is 
dominated at $M \approx 0$ 
by the magnetic form factor. The normalization at $M=0$, 
assuming the dominance of the magnetic transition, 
is sufficiently precise around $M=0$. With increasing $M$, the Coulomb form factor 
comes into play. In principle, parametrizations of the $\Delta(1232)$ Dalitz decay 
should be checked against the available photoproduction data. The quark counting rules 
constrain the extrapolation  to the $M \neq 0$ region.

In the present work, we apply the parametrizations of Ref. \cite{krivo02}, which 
are covariant and kinematically complete, i.e., formulated in terms of magnetic, 
electric, and Coulomb transition form factors. The $M$ dependence is based on the extended 
VMD (eVMD) model and constrained by photoproduction and electroproduction data for the 
form factors, by the transition helicity amplitudes of the nucleon resonances, and when available 
by the $\rho$- and $\omega$-meson decay branchings.

The authors of Ref. \cite{Bratkovskaya:2007jk} applied the parametrizations of Ernst \emph{et al.} \cite{Ern98}. The same parametrization has also been used for the PLUTO event 
generator of the HADES group \cite{Had06}. However, the work of Ernst 
\emph{et al.} \cite{Ern98} considered the magnetic form factor with no 
$M$ dependence, which  is a crude estimation. Furthermore, the 
kinematic factors of the $\Delta(1232)$ Dalitz decay in 
Refs.~\cite{Ern98,Had06,Bratkovskaya:2007jk} are incorrect, as discussed in 
Ref. \cite{Krivoruchenko:2001hs}.

Therefore, the 
interpretation of heavy-ion data still suffers from uncertainties 
unrelated to the complexity of heavy-ion dynamics; i.e., the various 
parametrizations of the resonance 
decays used as input in the transport models do not agree with each other.

\subsubsection{Collisional broadening}

\begin{figure}[!htb]
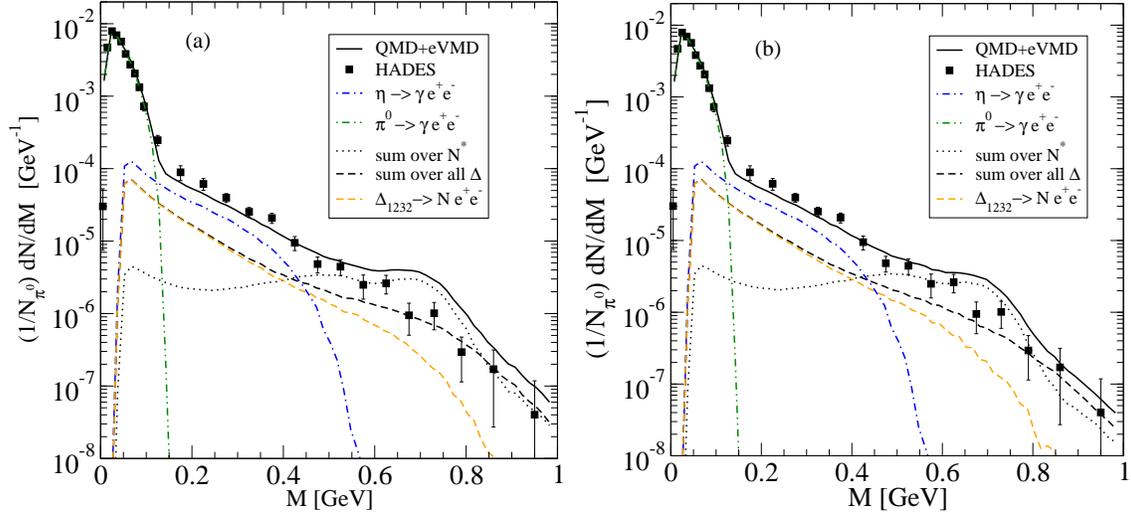

\centering
\includegraphics[width=.45\textwidth]{CC2.0_s.CB.r200om60_vacres.eps}
\includegraphics[width=.45\textwidth]{CC2.0_s.CB.r250om125_vacres.eps}
\caption{(Color online) Dilepton spectrum in C+C collisions at 2.0$A$ GeV for different values of the in-medium 
$\rho$ and $\omega$ widths. (a) $\Gamma_{\rho}^{\mathrm{tot}}(\rho_0)=200$ MeV and 
$\Gamma_{\omega}^{\mathrm{tot}}(\rho_0)=60$ MeV. (b) $\Gamma_{\rho}^{\mathrm{tot}}(\rho_0)=250$ MeV and
 $\Gamma_{\omega}^{\mathrm{tot}}(\rho_0)=125$ MeV.}
\label{CB} 
\end{figure}

Let us now turn to the introduction of in-medium effects
 according to the standard treatments and address first Fig.~\ref{CB}, 
in which the HADES data are compared with calculations where the 
possible broadening of the vector meson spectral function in medium is effectively 
taken into account through the introduction of a collisional 
width $\Gamma_V^{\mathrm{coll}}$. We present calculations that use 
a linear parametrization 
of the type $\Gamma_V^{\mathrm{tot}}(\rho_B)=
\Gamma_V^{\mathrm{vac}}+\rho_B/\rho_0\,\Gamma_V^{\mathrm{coll}}(\rho_0)$ 
to estimate 
the vector meson 
in-medium width $\Gamma_V^{\mathrm{tot}}(\rho_B)$. 

In a first approximation, we make no additional 
assumption concerning the energy dependence of the in-medium width; i.e., 
the same energy dependence is assigned to the collisional width as to 
the vacuum width~\cite{She03,Cozma06}.
In particular, the vector meson thresholds are kept the same 
as the vacuum ones, 
namely, $2m_{\pi}$ and  $3m_{\pi}$ for the $\rho$ and $\omega$ meson, 
respectively.
The approximation will be investigated below. 
Figure~\ref{CB}(a) refers to the assumption 
$\Gamma_{\rho}^{\mathrm{tot}}(\rho_0)=200$ and $\Gamma_{\omega}^{\mathrm{tot}}(\rho_0)=60$ MeV,
 which reflects the estimates 
of the CLAS and TAPS experiment for the collisional broadening of the 
$\rho$ and $\omega$ meson, respectively. Figure~\ref{CB}(b) refers to the 
assumption $\Gamma_{\rho}^{\mathrm{tot}}(\rho_0)=250$ ,
$\Gamma_{\omega}^{\mathrm{tot}}(\rho_0)=125$ MeV. 
The latter reflects the lower limit estimates emerged from the analysis
performed in Ref.~\cite{She03}, 
where the values of $300$ MeV and $200-300$ MeV, respectly, for the $\rho$ and $\omega$ meson widths at an average
density of $1.5\rho_0$ were extracted from fits to the DLS data.

We observe a suppression of the peak with respect to the vacuum case, 
more pronounced in case Fig.~\ref{CB}(b) than in Fig.~\ref{CB}(a). 
However, in both cases, the experimental data 
are still overestimated around $M\sim0.7$ GeV, mainly due to the 
still significant contribution of the $N^*(1535)$ resonance. 
The Dalitz decay of this resonance plays a dominant role in the 
determination of the dilepton spectrum in the region around the 
vector meson peak, due to its strong coupling to the $\omega$ meson. 
On the one hand, the HADES data seem to favor a 
smaller contribution of the 
$N^*(1535)$ resonance in the peak region; 
on the other hand, however,  
dilepton production data in $p$+$p$ collision have been well 
described under the same assumptions for the 
coupling to the $N^*$(1535). 
This shows that the 
contribution of the $N^*(1535)$ Dalitz decay, significant in 
elementary reactions and thus in vacuum, is 
partially reduced in heavy-ion collisions thanks 
to in-medium effects. We conclude that the HADES 
data suggest a stronger 
in-medium modification of the $\omega$ properties than the 
one taken into account in Fig.~\ref{CB}.

Let us now investigate the effect 
of different choices for the energy dependence of the collisional width.  
In this context, we would like to point out that the
mere fact of having, and facing, a certain freedom in the choice of an energy
dependence of the collisional width shows exemplarily the limits 
that such schematic models carry.
Such choices can be based on more or less educated guesses. 
However, if microscopic calculations of in-medium effects are performed, 
 energy dependences are fixed from theory, which 
should be fulfilled as a minimal requirement for a consistent investigation
  of  vector meson in-medium properties.
Obviously a microscopic calculation of the exact energy dependence of the 
collisional broadening is equivalent to a full model 
calculation of the in-medium spectral function. This well be done later on in
this work. 

For the moment, we investigate the consequences of various 
approximations on a schematic level. 
For this purpose, we extract possible energy dependences 
of the collisional widths on the basis of qualitative considerations and 
consider the influence on the shape of the final dilepton spectrum. 
Schematically, the collisional broadening that a vector meson 
acquires is attributed to an absorption process of the type 
$V+N\rightarrow R \rightarrow \pi+N$. To simplify, we approximate the 
corresponding 
phase space by the phase space for the process 
$M+m_N \rightarrow m_\pi+m_N$ and assume that 
the resonance decay proceeds through a $p$ wave. This latter freedom demonstrates 
again the limits of such  schematic procedures. However, since this estimate is 
only qualitative, let us neglect for the moment these refinements.

\begin{figure}[!htb]
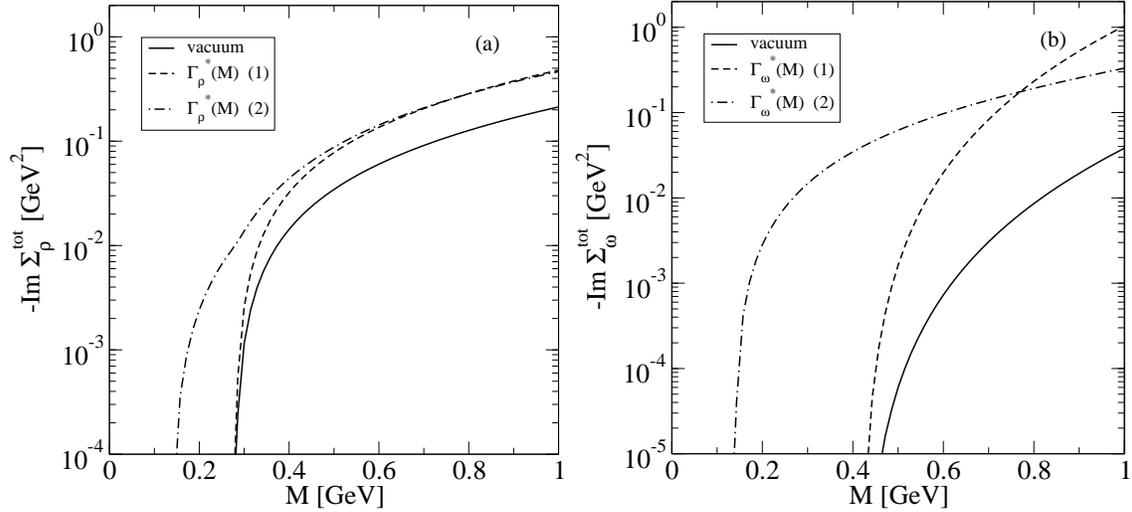

\centering
\includegraphics[width=.45\textwidth]{comp_im_rho.eps}
\includegraphics[width=.45\textwidth]{comp_im_ome.eps}
\caption{(a) Imaginary part of the in-medium self-energy 
$-\Im\Sigma_\rho^{\mathrm{tot}}(\rho,M)=m_\rho \Gamma_{\rho}^{\mathrm{tot}}(\rho,M)$ 
of the $\rho$ meson in vacuum (full line) and at $\rho=2\,\rho_0$ for  
$\Gamma_{\rho}^{\mathrm{tot}}(\rho_0, m_\rho)=250$ MeV 
(dashed and dashed-dotted lines). (b) 
Imaginary part of the in-medium self-energy 
$-\Im\Sigma_\omega^{\mathrm{tot}}(\rho,M)=m_\omega \Gamma_{\omega}^{\mathrm{tot}}(\rho,M)$ 
of the $\omega$ meson in vacuum (full line) and at $\rho=2\,\rho_0$ for
 $\Gamma_{\omega}^{\mathrm{tot}}(\rho_0,m_\omega)=125$ MeV(dashed and 
dashed-dotted lines). For both panels, the dashed line corresponds to 
the assumption 
that the collisional width has the same energy dependence as the 
vacuum width.
The dashed-dotted line corresponds to the assumption that the 
collisional width 
has the energy dependence in  Eq.~(\ref{gammacoll_phs}).}
\label{ims}
\end{figure}

One obtains
\beq
 \Gamma_V^{\mathrm{coll}}(M,\rho)=\Gamma_V^{\mathrm{coll}}(m_V,\rho)\,
 \left( \frac{m_V+m_N}{M+m_N} \right ) 
 \left( \frac{q(M+m_N,m_N,m_\pi)}{q(m_V+m_N,m_N,m_\pi)}\right )^3~,
 \label{gammacoll_phs}
 \eeq
 with
 \beq
 q(M+m_N,m_N,m_\pi)=\frac{\sqrt{[(M+m_N)^2-(m_N+m_\pi)^2]
[(M+m_N)^2-(m_N-m_\pi)^2]}}{2\,(M+m_N)}~.
 \eeq
As one can see, in this approximation, the vector meson threshold is 
shifted from $2 m_\pi$ to $m_\pi$ for the $\rho$ meson and from $3 m_\pi$ 
to $m_\pi$ for the 
$\omega$ meson. The choice affects the shape of the $\omega$ width much 
more than the shape of the $\rho$ width.
The influence of the choice for the energy 
dependence of the collisional width is 
illustrated in 
Fig.~\ref{ims} for the case $\rho_B=2\rho_0$ 
and for $\Gamma_{\rho}^{\mathrm{tot}}(\rho_0, m_\rho)=250$ and 
$\Gamma_{\omega}^{\mathrm{tot}}(\rho_0,m_\omega)=125$ MeV. In particular, 
for the $\omega$ meson, the shift of the 
threshold leads to a large enhancement of the $\omega$ width at lower 
invariant masses.

However, one has to keep in mind  that the 
in-medium $\rho$ and $\omega$ widths enter into the expressions for the 
the covariant form factors, see Eqs.~(\ref{formfac})--(\ref{inmedampl}).
Their  modulus squared determines the 
width $\Gamma (R\rightarrow N\gamma^{*})$ [Eq.~(\ref{OK!})]. 
Thus, only when appreciable 
differences arise in the covariant form factors will 
the difference in the energy dependence of the 
in-medium vector meson width be visible in the final dilepton spectrum. 
Now let $\Gamma_\rho^{* [1]} (M)$ be 
the $\rho$ meson in-medium width with an energy dependence analogous 
to the vacuum 
width and $\Gamma_\rho^{* [2]} (M)$ the  $\rho$ meson in-medium width with an energy dependence 
according to Eq.~(\ref{gammacoll_phs}).
Correspondingly, we set
\beqa
F_\rho^{[1/2]}&=&\frac{m_\rho^2}{m_\rho^2- i m_\rho \Gamma_\rho^{* [1/2]} (M)-M^2 }~,\\  
F_\omega^{[1/2]}&=&\frac{m_\omega^2}{m_\omega^2- i m_\omega \Gamma_\omega^{* [1/2]} (M)-M^2 }~.  
\eeqa
We refer now to the $\omega$ meson, but the same considerations are 
valid for the $\rho$ meson. 
It can be easily realized that $|F_\omega^{[2]}|^2\approx |F_\omega^{[1]}|^2 \approx 1$ when 
$(m_\omega^2-M^2)^2 \gg (m_\omega \Gamma_\omega^{* [i]})^2$ ($i=1,2$). 
Thus, the mass region where sensible differences between the two cases can be found  
is typically restricted to the mass region around the 
vector meson peak. Concerning the $\rho$ meson, $\Gamma_\rho^{* [2]}$ and 
$\Gamma_\rho^{* [1]}$ are practically identical in the region 
of the vector meson peak, as can be seen from Fig.\ref{ims}. 

Therefore, we do not expect differences between $|F_\rho^{[2]}|^2$ and 
$|F_\rho^{[1]}|^2$. Concerning  the $\omega$ meson, $\Gamma_\omega^{* [2]}$  
and $\Gamma_\omega^{* [1]}$ differ substantially in the peak region, although
the main differences arise at lower masses, i.e., 
from slightly above $m_\pi$ up to slightly above $3m_\pi$, 
because of the different thresholds. 
In addition, one should  also consider interference terms 
of the form  
$F_\rho^{[i]}F_\omega^{[i]}$. These terms
 can, however, drive either a constructive or destructive interference, 
and therefore 
it is not possible to comment on their effect in general 
within a simple scheme. 

\begin{figure}[!htb]
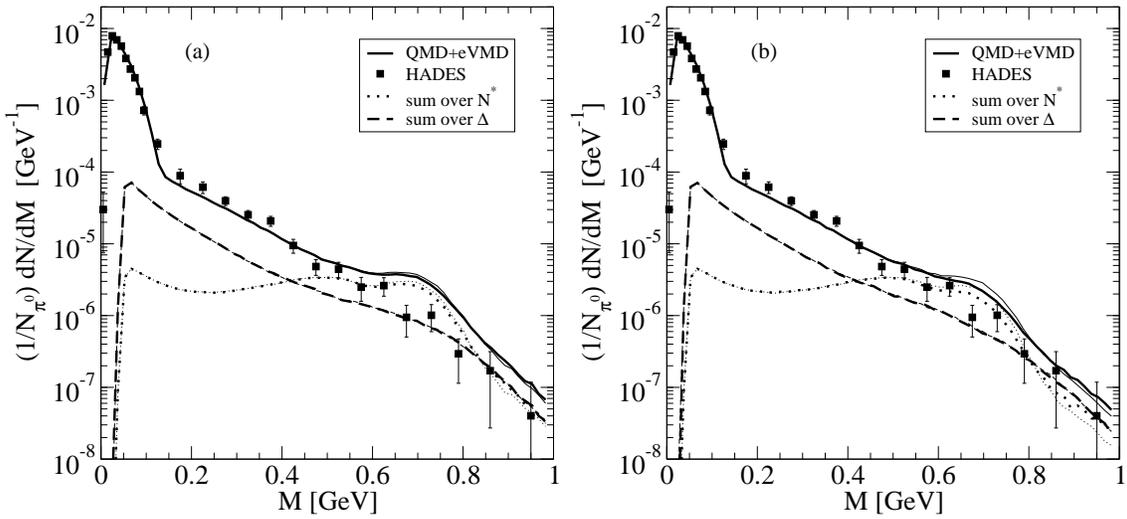

 \centering
 \includegraphics[width=.45\textwidth]{CC2.0_s.CB.r200om60_vacres_comp.eps}
 \includegraphics[width=.45\textwidth]{CC2.0_s.CB.r250om125_vacres_comp.eps}
 \caption{Dilepton spectrum in C+C collisions at 2.0$A$ GeV for different 
 values of the in-medium 
 $\rho$ and $\omega$ widths and different choices for the energy dependence 
 of the collision width. 
 The thick lines refer to an energy dependence  estimated from the 
 $V+N\rightarrow R \rightarrow \pi+N$ as discussed in the text. The thin lines
  correspond to the same calculations shown in Fig.~\ref{CB} and are shown 
 for comparison.
 (a) $\Gamma_{\rho}^{\mathrm{tot}}(\rho_0)=200$ and 
 $\Gamma_{\omega}^{\mathrm{tot}}(\rho_0)=60$ MeV. 
(b) $\Gamma_{\rho}^{\mathrm{tot}}(\rho_0)=250$ and
  $\Gamma_{\omega}^{\mathrm{tot}}(\rho_0)=125$ MeV.}
 \label{CB_phsp}
 \end{figure}

The resulting  dilepton spectra obtained for the two choices discussed above 
are shown in Fig.~\ref{CB_phsp}.  
Here one finds  that the contributions from the $\Delta$ resonances, 
which couple only to the $\rho$ meson, are practically identical 
in the two cases. Slight differences are 
visible for the $N^*$
 resonances around the vector meson peak. The differences are more evident 
in the case of larger values of the widths, Fig.~\ref{CB_phsp}(b). 
However, even in this case, the total spectra differ at most by a 
factor of 1.3~\footnote{
Here we refer to the maximum value of the ratio of the two spectra.}.

For a consistent 
evaluation of the energy dependence resulting from 
$V+N\rightarrow R \rightarrow \pi+N$ processes, 
one should sum up over all important resonances that couple 
to the $N+V$ system, each taken with a different weight according to 
their relative coupling strength, and 
determine for each mode the corresponding 
angular momentum of the $\pi N$ scattering amplitude. 
Moreover, the invariant mass squared of the intermediate resonance 
would be $s=(p_N+p)^2$ which leads to a dependence on the three-momentum 
$\mathbf{p}$ of the vector meson. 
It is then clear that such a procedure would finally be 
analogous to the evaluation of the full spectral functions. 
In fact, the $V+N\rightarrow R \rightarrow \pi+N$ channel is one of the processes 
consistently 
included in our calculation of the spectral functions, since the $N\pi$ 
channel 
is one of the channels entering in the expression of the total width of the 
resonance.

To conclude, already these first estimates based on the collisional 
broadening scenario demonstrate that the HADES data show clear evidence 
for a strong in-medium
modification of the vector meson properties. 
Figure \ref{CB_phsp} demonstrates, on the other hand,  that 
though the two different choices under discussion 
lead to significant deviations of the vector meson widths, particularly 
concerning their threshold behavior, such effects are washed out to a 
large extent 
in the final spectra. However, the same 
argument demands a theoretical description that is as precise 
as possible; i.e.,  realistic spectral 
functions should be applied rather than pushing schematic models 
too far.

\subsubsection{Dropping mass scenario}

\begin{figure}[!htb]
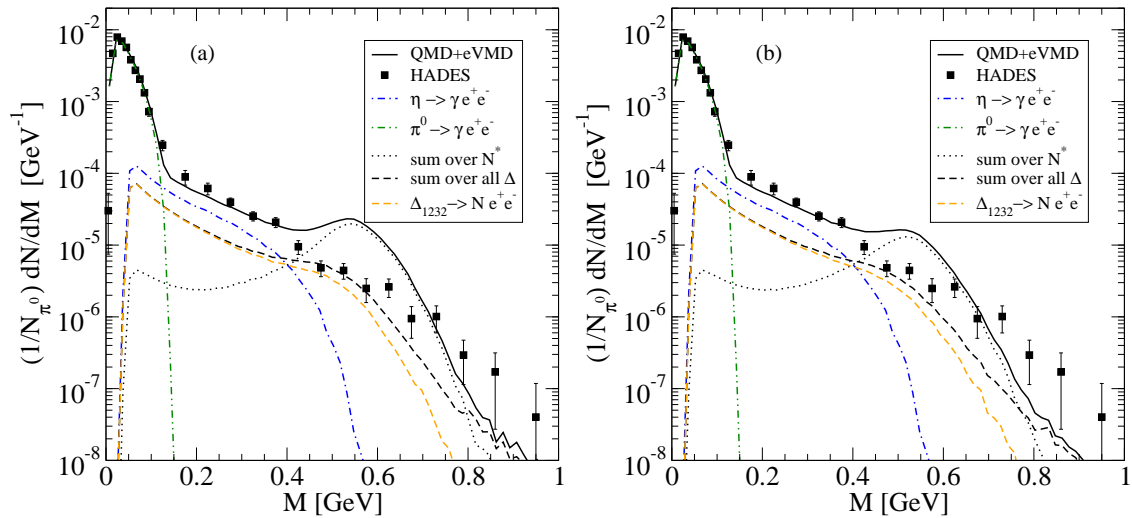

\centering
\includegraphics[width=.45\textwidth]{CC2.0_s.BR.r200om60_vacres.eps}
\includegraphics[width=.45\textwidth]{CC2.0_s.BR.r250om125_vacres.eps}
\caption{(Color online) Dilepton spectrum in C+C collisions at 2.0 AGeV 
for 
different values of the in-medium 
$\rho$ and $\omega$ widths when an in-medium vector meson mass
 $m_V^* = m_V(1- \alpha\rho_B/\rho_0)$ is introduced. 
(a) $\Gamma_{\rho}^{\mathrm{tot}}(\rho_0)=200$ and 
$\Gamma_{\omega}^{\mathrm{tot}}(\rho_0)=60$ MeV. (b) $\Gamma_{\rho}^{\mathrm{tot}}(\rho_0)=250$ and
 $\Gamma_{\omega}^{\mathrm{tot}}(\rho_0)=125$ MeV.}
\label{BR}
\end{figure}

However, before adopting realistic spectral functions, we 
want to investigate the  dropping mass 
scenario \emph{\`a la} Brown-Rho, which has been widely used in the 
literature. Thus, we performed 
calculations for an in-medium scenario that differs from the previous 
one by the
additional assumption that the vector meson mass 
scales with density 
according to a $m_V^* = m_V (1- \alpha\rho_B/\rho_0)$ law, with $\alpha=0.2$.
 The results are 
shown in Fig.~\ref{BR}, where Fig.~\ref{BR}(a) refers to the choice 
$\Gamma_{\rho}^{\mathrm{tot}}(\rho_0)=200$ MeV and $\Gamma_{\omega}^{\mathrm{tot}}(\rho_0)=60$ MeV, 
and  Fig.~\ref{BR}(b) to 
the choice $\Gamma_{\rho}^{\mathrm{tot}}(\rho_0)=250$, 
$\Gamma_{\omega}^{\mathrm{tot}}(\rho_0)=125$ MeV.
The inclusion of a dropping in-medium vector meson mass results in a global shift of the vector meson 
spectral strength to lower masses. 
Thus, the corresponding theoretical spectrum is enhanced at lower invariant
masses resulting in a sizable overestimation of the experimental data in the $0.4\leq M
\leq 0.7$ GeV mass region. 
At the same time, the experimental data 
are underestimated in the region around and above the vector meson peak 
because of the lack of spectral strength around the 
(vacuum) vector meson peak. 
Note that the same underestimation of the vector meson peak 
was observed when adopting the 
dropping mass scenario to the recent high resolution CERES data~\cite{Adamova:2006nu}.
The CERES analysis focused, however, only on the in-medium $\rho$ meson. 

Concerning the low mass region, $m_\pi\leq M \leq 0.4$ GeV, the presence of
additional strength moves the spectrum closer to the experimental data in 
the mass region $M\sim0.3$--$0.4$ GeV. 
However, this region  remains 
slightly but systematically underestimated. In summary, one can conclude that
a naive Brown-Rho scaling is too schematic in order to explain 
the spectrum. This finding is consistent with the previous 
theoretical analysis of the DLS data at 1$A$ GeV~\cite{BCRW98,Ern98}.

\subsubsection{In-medium spectral functions}

\begin{figure}[!htb]
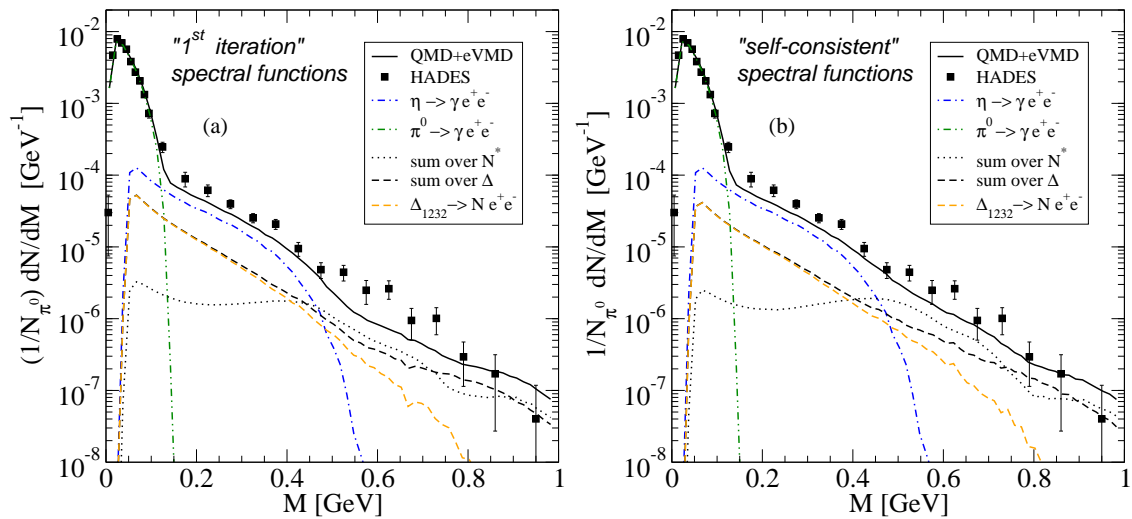

\centering
\includegraphics[width=.45\textwidth]{CC2.0_s.sf_nonres_vacres.eps}
\includegraphics[width=.45\textwidth]{CC2.0_s.sf_selfcon_vacres.eps}
\caption{(Color online) Dilepton spectrum in C+C collisions at 2.0$A$ GeV 
resulting from 
the inclusion of $\rho$- and $\omega$-meson spectral functions calculated 
within the NRD+eVMD model. The spectral functions affect 
the branching ratios for the Dalitz 
decays of the baryon resonances, as explained in the text.
(a) Inclusion of 
vector meson self-energies determined 
from vacuum nucleon resonance properties. 
(b) Inclusion of 
vector meson self-energies calculated 
in a self-consistent iteration scheme 
that takes into account the in-medium modifications of the 
nucleon resonance widths induced 
by the in-medium spectral functions of 
the vector mesons.
}
\label{spf}
\end{figure}

Let us now pass to the investigation of in-medium properties 
based on the in-medium self-energies of the vector 
mesons calculated within NRD+eVMD.
First, we present in Fig.~\ref{spf}(a) the dilepton spectrum 
obtained with ``first iteration'' $\rho$ and $\omega$ spectral functions, 
i.e., neglecting the in-medium modification of the nucleon resonance widths. 

\begin{figure}[!htb]
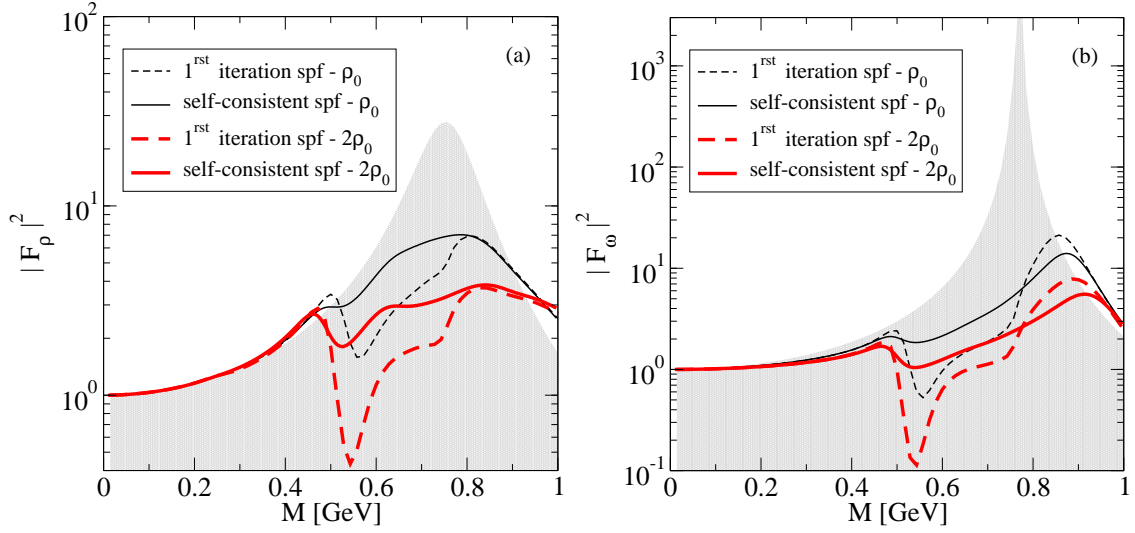

\centering
\includegraphics[width=.45\textwidth]{comp_ff_rho.spf.eps}
\includegraphics[width=.45\textwidth]{comp_ff_ome.spf.eps}
\caption{(Color online) (a) Modulus squared of the $\rho$ meson contribution 
to the covariant form factor $|F_\rho|^2$ 
at $\rho_B=\,\rho_0$ (thin lines) and $\rho_B=2\,\rho_0$ (thick lines). 
(b) Modulus squared of the $\omega$ meson contribution to 
the covariant form factor $|F_\omega|^2$ 
at $\rho_B=\,\rho_0$ (thin lines) and $\rho_B=2\,\rho_0$ (thick lines). 
For both panels the dashed lines correspond to vector meson 
self-energies calculated 
from vacuum nucleon resonance properties. The full lines correspond to 
vector meson self-energies calculated 
in a self-consistent iteration scheme that takes into account the in-medium 
modifications of the nucleon resonance widths induced 
by the in-medium spectral functions of the vector mesons.
Shaded areas indicate the  vacuum form factors.
}
\label{comp_ff.spf}
\end{figure}

The spectral functions induce a depletion of the theoretical spectrum in the mass 
region $0.45\leq M \leq 0.75$ GeV which is not supported by the data. 
The result can be better understood with the help of Fig.~\ref{comp_ff.spf}, 
which shows the corresponding $\rho$ and $\omega$ contributions 
which enter into the nucleon resonance form factors and determine thus 
the dilepton production rates. The form factors are determined 
at saturation density and twice  saturation density, 
$\rho_B=\rho_0$ and $\rho_B=2\rho_0$, in both cases for a 
vector meson at rest relative to the nuclear medium (dashed lines).

The complex structure of the mesonic self-energies is clearly reflected 
in the form factors, which no longer preserve the simple 
Lorentzian-like shape typical for the vacuum. 
In particular, for the $\rho$ as well as for the $\omega$ 
we observe a strong minimum around  
$0.5\lesssim M \lesssim 0.6$ GeV between two 
maxima at $0.4\lesssim M \lesssim 0.5$ GeV and $M\sim 0.8$ GeV.

\begin{figure}[!htb]
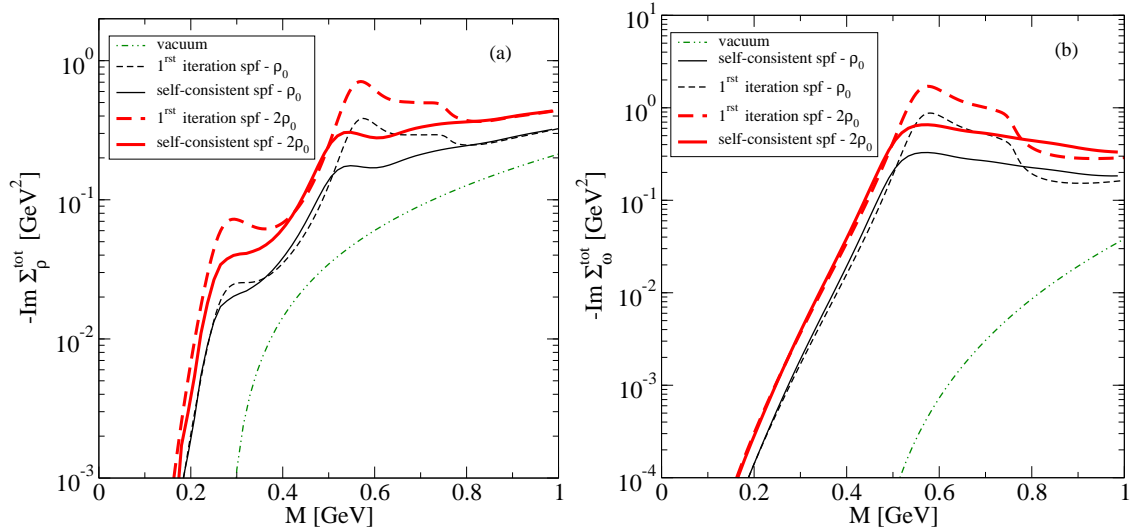

\centering
\includegraphics[width=.45\textwidth]{comp_im_rho.spf.eps}
\includegraphics[width=.45\textwidth]{comp_im_ome.spf.eps}
\caption{(Color online) Imaginary part of the in-medium self-energy 
of the (a) $\rho$ meson and (b)  $\omega$ meson, 
in vacuum (dashed-double-dotted lines), at $\rho_B=\,\rho_0$ 
(thin lines) and $\rho_B=2\,\rho_0$ (thick lines). 
Dashed lines correspond to vector meson self-energies calculated 
from vacuum nucleon resonance properties. Full lines correspond to vector meson self-energies calculated 
in a self-consistent iteration scheme that takes into account the in-medium modifications of the nucleon resonance widths induced 
by the in-medium spectral functions of the vector mesons.}
\label{comp_im.spf}
\end{figure}

The particular shape of the form factor is determined by the interplay 
of both the real and imaginary part of the self-energy. 
However, switching off the real part of the self-energy, 
we observe that the depletion of the form factor between $M\sim0.5$ and 
$0.8$ GeV 
is mainly caused by large values of the imaginary part of the 
self-energy in this region. The latter is shown in Fig.~\ref{comp_im.spf}. 
This increase is due to the strong coupling to specific resonances, 
i.e., the $N^*(1520)$ for the $\rho$ meson and the $N^*(1535)$ for the $\omega$ meson. 
The corresponding 
bump structure is a typical feature for this class of models which couple  
vector mesons to resonance-hole states. 

The inclusion of the in-medium resonance properties caused by 
the vector meson spectral functions, i.e., self-consistency, reduces 
the imaginary part of the self-energy in this region 
(see Fig.~\ref{comp_im.spf}). In the case of the $\omega$ meson, 
for example, the reduction at $M=0.57$ GeV is about a factor of 2.5. 
As a consequence, the form factors, shown in  
Fig.~\ref{comp_ff.spf}, are enhanced.
 
This has an effect on the dilepton spectrum. 
The spectrum obtained with self-consistent 
spectral functions is shown in Fig.~\ref{spf}(b).  
The inclusion of the in-medium properties of the nucleon resonances 
moves the theoretical spectrum closer to the experimental data in the mass 
region $0.45\leq M \leq 0.75$ GeV. This demonstrates the importance of higher order effects, i.e.,
taking in-medium modifications for the nucleon 
resonances into account when the vector meson properties are described by 
the coupling to nucleon-resonance hole states.

For the mass region $M>0.4$, we conclude that 
the parameter-free 
determination of the in-medium dilepton spectrum, 
performed within an approach that attempts to  describe 
\emph{simultaneously and with the same model parameters} \footnote{This 
statement refers to the resonance+eVMD model parameters. The 
additional parameters of the RQMD transport model (cross sections, potentials, 
etc.) 
have not been changed.} the phenomena of 
dilepton and vector meson production as well as their in-medium 
modifications, gives a reasonable description of the experimental data.

However, some data points remain still underestimated. This suggests that 
the NRD+eVMD model predicts a too strong absorption of vector mesons. 
One possible reason for the present underestimation 
of the experimental data is the use of some poorly 
constrained eVMD model parameters, 
in particular the 
$RN\omega$ couplings. Probably the most relevant case is the 
$N^*(1535)$ resonance, with its strong coupling to the $\omega$ meson 
predicted by the eVMD model though a decay of this resonance 
to $N\omega$ has not been measured yet. 
Another reason might be that the $\omega$-meson spectral 
function in particular is not normalized in the mass region 
of our interest. The violation of normalization ranges from 
about 30\% at $\rho=\rho_0$ to about 45\% at $\rho=2\rho_0$.\footnote{
Integral evaluated in 
the mass region up to $1.5$ GeV.} 
In principle, this represents no inconsistency, since spectral functions 
must satisfy the sum rule in the entire invariant mass range 
(up to $M=\infty$) and not 
necessarily already in the finite mass interval in which we work. 
This effect should, however, be investigated in future work.


Regarding the low mass region, $m_\pi\leq M \leq 0.4$ GeV, the introduction 
of in-medium spectral functions does not provide a solution for the 
underestimation of the experimental data. On the contrary, because of the 
finite value of the imaginary part of the self-energy at 
$M\sim 0$ for high vector meson three-momenta  $\mathbf{p}$   
[$\Im\Sigma_V^{\mathrm{tot}} (M=0) \neq 0$ for $\mathbf{p} \neq 0$], at high momenta we have 
$|F_V (M=0)|^2<1$ with a consequent reduction of strength.
We can therefore conclude that for the explanation of the low mass region, one 
has to take into account additional effects and/or sources.

\section{Conclusions}

We  determined the modification of the $\rho$ and $\omega$ meson properties 
in nuclear matter  within a resonance model and investigated the 
nonresonant contributions to the vector meson self-energy.
For both vector mesons, we found a substantial broadening of the width and a 
significant shift of spectral strength down to smaller invariant masses. 
In particular at small momenta, the coupling of the $\rho$ meson to the 
$N^*(1520)N^{-1}$ state and that of the $\omega$ meson to the $N^*(1535)N^{-1}$ 
state lead to pronounced double-peak structures in the spectral functions. 
In a first approximation, the spectral functions 
were determined from vacuum 
nucleon resonance properties. 
Going beyond this approximation, the in-medium 
modification of the nucleon resonance widths 
induced by the modified $\rho$ and $\omega$ mesons has been included. 
This leads to a self-consistent calculation of the vector meson 
spectral functions, which mainly 
reduces the peaks due to the coupling to $N^*(1520)N^{-1}$ and 
$N^*(1535)N^{-1}$ states.

In a next step, we investigated the influence of different 
in-medium scenarios for 
the vector mesons on the dilepton production rate in 
heavy-ion collisions. 
The dilepton spectrum has been calculated exemplarily for the reaction 
C+C  at 2.0$A$ GeV for 
which experimental data have been recently 
released by the HADES Collaboration.

Already the estimates based on a schematic collisional broadening 
scenario, i.e., the comparison with data, support strong in-medium modification 
of the vector meson properties. In  the dropping mass scenario, we found, 
even when taking additionally into account the collisional broadening 
of the vector meson widths, that the dilepton spectrum 
overestimates the experimental data at invariant masses below 
the vector meson peak and 
underestimates them in the region around and above 
the peak.

Finally, we went beyond the schematic inclusion of in-medium effects 
and included the vector meson properties consistently, 
i.e., in terms of the in-medium self-energies microscopically 
calculated within our model. 
We found that self-energies determined from vacuum nucleon resonance 
properties give a poor description of the experimental data in the invariant 
mass region $0.45\leq M \leq 0.75$ GeV.
On the contrary, the self-consistent 
iteration scheme provides a reasonable 
description of the data in the same mass region. 
This demonstrates the importance of consistent inclusion of in-medium 
properties. 

However, for the low mass region ($m_\pi\leq M \leq 0.4$ GeV)
we found that the 
inclusion of $\rho$ and $\omega$ spectral functions does 
not improve the theoretical description of  the dilepton spectrum
and experimental data remain slightly underestimated.

In summary, the investigation represents a first step toward a 
unified understanding of dilepton spectra and vector meson properties 
in heavy-ion collisions at intermediate energies. 
The same model and the same set of parameters were used to 
describe the 
interconnected phenomena of dilepton and vector meson production 
and their in-medium modifications.
Forthcoming data, from elementary reactions as well as from 
heavy-ion collisions of heavy systems, 
will certainly help to further reduce 
still existing model uncertainties.
 An extension  of the present approach to finite 
baryon chemical potential and temperature would further 
allow one to test the spectral properties beyond HADES conditions, 
e.g., by a comparision with the NA60 data.

\section*{Acknowledgments}
We are grateful to the HADES Collaboration for help concerning 
the HADES filter program and for providing us with the experimental data.
This work was supported by the European Graduate School 
Basel-Graz-T\"ubingen 
and by the RFBR Grant No. 06-02-04004 and the DFG Grant 
No. 436 RUS 113/72/0-2.


\appendix

\section{Gauge invariance in eVMD}
\setcounter{equation}{0} 

The VMD model and its modifications introduce the mixing of a photon with 
vector mesons $\rho^{0}$, $\omega$, $\phi$, etc. Such a mixing can, 
in principle, generate finite photon masses and destroy gauge invariance. This problem has been 
solved for the VMD model by Kroll, Lee and Zumomino \cite{KROLL} 
constructing an effective Lagrangian for photons and vector mesons 
which reproduces the VMD predictions. We present first a distinct consistency proof and then show how 
the method \cite{KROLL} can be generalized to the eVMD model.

\subsection{Final-state interaction (FSI)  method}
We start from an effective Lagrangian involving pions interacting with 
photons. An example of such a Lagrangian is the nonlinear $\sigma$ model 
and, more generally, the chiral perturbation theory (ChPT) to a fixed order of the loop expansion. The
vector mesons appear as resonances in the two-pion scattering channel 
($\rho$ mesons)
and the three-meson scattering channel ($\omega$ mesons). 


\begin{figure}[!htb]
\begin{center}
\includegraphics[angle=0,width=5cm]{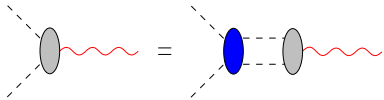}
\caption{
(Color online) Diagrammatic representation of the FSI of pions (dashed lines) contributing to the
form factor in the $\rho$-meson channel. 
The photon line is shown as a wavy line.}
\label{fig1}
\end{center}
\end{figure}

Let us consider an absorption of a photon in an isovector channel, 
as shown in Fig. \ref{fig1}.
Applying two-body unitarity and taking into account 
analyticity (see, e.g., Ref. \cite{BD}, Chap. 18), we replace the 
point-like vertex $e$ by $eP_{l}(t)/D_{J}(t)$ where $t=q^2$ 
is the photon momentum squared, $P_{l}(t)$ is a 
polynomial of the degree $l$, and $D_{J}(t)$ is the Jost function 
defined in terms of the $p$-wave isovector two-pion scattering phase shift $\delta(t)$:
\begin{equation}
D_{J}(t) = \exp\left[-\frac{t}{\pi}\int_{t_{0}}^{\infty}\frac{\delta(t^{\prime})dt^{\prime}}{t^{\prime}(t - t^{\prime})}\right]~,
\label{JOST}
\end{equation}
where $t_{0}$ is the two-pion threshold.

In the no-width approximation, the phase shift accounting 
for the existence of $n$ resonances is given by
\begin{equation}
\delta(t) = \sum_{k=1}^{n}\pi \theta(t - m^2_{k})~,
\label{J2}
\end{equation}
where $m_{k}$ is the mass of the $k$th radial excitation of the $\rho^{0}$ meson. 
Substituting this expression into Eq.(\ref{JOST}), we obtain
\begin{equation}
F(t) = P_{l}(t)\prod_{k=1}^{n}\frac{m^2_{k}}{t - m^2_{k}}.
\label{J3}
\end{equation}
The requirement $F(t) \rightarrow 0$ at $t \rightarrow \infty$ gives $l<n$. 

Analytical functions are fixed by their singularities. 
The representation (\ref{J3}) can be rewritten in an equivalent additive form
\begin{eqnarray}
F(t) = \sum_{k=1}^{n}c^{k}\frac{m^2_{k}}{m^2_{k} - t}~,
\label{DE3}
\end{eqnarray}
where $c^{k}$ are some coefficients. The normalization 
condition $F(0) = 1$ and the quark counting rules impose constraints for $c^{k}$. 

The effective pion Lagrangian is well defined,  since pions are 
stable particles which exist as asymptotic states. In the approach 
presented above, the problem of gauge invariance does not appear, 
since gauge invariance of the effective Lagrangian ensures a transverse 
polarization tensor of photons and the vanishing photon mass. The vector mesons
are resonances accounted for by the the final-state interactions. 

\subsection{Effective Lagrangian method}

The vector mesons are unstable particles and do not exist as asymptotic 
states. Nevertheless, the effective Lagrangian method is useful in 
formulating 
vector meson effective interactions. 
Kroll, Lee, and Zumomino \cite{KROLL} proposed an effective 
Lagrangian for the VMD model to illustrate its gauge invariance. We extend their 
arguments for a family of $n$ 
$\rho^{0}$ mesons interacting with photons. An effective 
Lagrangian can be written as
\begin{equation}
\mathcal{L}_{eff} = - \frac{1}{4}              F_{\mu \nu}    F_{\mu \nu} 
                  + \sum_{k=1}^{n}\left( - \frac{1}{4} G_{\mu \nu}^{k}G_{\mu \nu}^{k}
+ \frac{1}{2} m^2_{k}B_{\mu}^{k}B_{\mu}^{k} + \frac{e}{2g^{k}} G_{\mu \nu}^{k} F_{\mu \nu} \right) - \left( eA_{\mu} + \sum_{k=1}^{n} h^{k}B_{\mu}^{k} \right)J_{\mu}~, 
\label{EFFLAG}
\end{equation}
where $A_{\mu}$ is the electromagnetic vector potential, $B_{\mu}^{k}$
is the $k$th $\rho^{0}$-meson vector potential, $J_{\mu}$ is a hadron conserved current,
and $F_{\mu \nu} = \partial_{\nu} A_{\mu} - \partial_{\mu} A_{\nu}$ and 
$G_{\mu \nu}^{k} = \partial_{\nu} B_{\mu}^{k} - \partial_{\mu} B_{\nu}^{k}$.

Lagrangian (\ref{EFFLAG}) is gauge invariant with respect to gauge 
transformations of the electromagnetic vector potential, so the photon 
interactions with the vector mesons do not violate gauge invariance and, 
in particular, do not generate a photon mass. 

\begin{figure}[!htb]
\begin{center}
\includegraphics[angle=0,width=7.37cm]{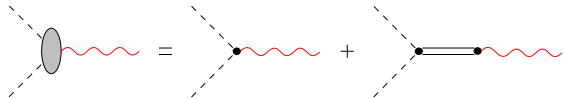}
\caption{
(Color online) Diagrammatic representation of the photon interaction with the electromagnetic current 
using the effective Lagrangian couplings. The first diagram shows the direct photon 
coupling and the second one shows the coupling through the family of $n$ 
$\rho$ mesons (double solid line).
The photon line is shown as a wavy line.}
\label{fig2}
\end{center}
\end{figure}

It remains to be shown that the coupling constants $g^{k}$ and $h^{k}$ 
can be chosen such that they reproduce the eVMD predictions. 
For each vector meson, we consider the two diagrams 
shown on Fig. \ref{fig2}. Their sum gives 
\begin{eqnarray}
F(t) = 1 + \sum_{k=1}^{n}\frac{1}{g^{k}}\frac{t}{m^2_{k} - t}h^{k}.
\label{DE}
\end{eqnarray}
The spectral functions of the form factors and their 
asymptotic behavior depend on the type of transition. 
The usual VMD appears for $n=1$. It corresponds to 
asymptotics $F(t) \sim 1/t$ at $t \rightarrow \infty$. 
If we set $h^{1} = g^{1}$, the monopole form factor is reproduced:
\begin{eqnarray}
F(t) = \frac{m^2_{1}}{m^2_{1} - t}.
\label{DE}
\end{eqnarray}
For $m_{1} = m_{\rho}$ it describes well the pion form factor in the space-like region.
The model \cite{FF,GS} of the pion form factor, which represents an 
improvement of the VMD to account for the analyticity and two-body 
unitarity of the pion form factor, and the $\rho$-meson width, works 
well in both the space- and time-like regions. 


In the case of eVMD, we set $h^{k} = c^{k}g^{k}$, where $c^{k}$ are 
some coefficients. The form factor $F(t)$ should decay at infinity, 
so we obtain
\begin{eqnarray}
\sum_{k=1}^{n}c^{k} = 1.
\label{DE2}
\end{eqnarray}
Then the usual representation (\ref{DE3}) of the eVMD form factors follows.

The quark counting rules can be 
satisfied selecting the coefficients $c^{k}$. $F(t) \sim 1/t^2$
gives
\begin{eqnarray}
\sum_{k=1}^{n}c^{k}m^2_{k} = 0.
\label{DE4}
\end{eqnarray}
Equations (\ref{DE2}) and (\ref{DE4}) have a unique solution for $n=2$.
$F(t) \sim 1/t^3$ requires the existence of at least $n=3$ vector mesons 
and so on.

\end{document}